\documentclass[twocolumn]{article}
\usepackage[a4paper,margin=0.95in]{geometry}  
\usepackage{microtype}
\usepackage{graphicx}
\usepackage{subfigure}
\usepackage{siunitx}
\usepackage{booktabs} 
\usepackage{caption}

\usepackage{hyperref}

\usepackage{amsmath}
\usepackage{amssymb}
\usepackage{mathtools}
\usepackage{amsthm}
\usepackage[numbers]{natbib}  
\usepackage{titlesec}  
\usepackage{fancyhdr}  

\usepackage[capitalize,noabbrev]{cleveref}

\usepackage{xcolor}
\usepackage{pgfplots}
\pgfplotsset{compat=1.18} 
\usepackage{tikz}
\usetikzlibrary{spy}

\tikzset{
    img/.style={inner sep=0,anchor=center},
    mylabel/.style={anchor=north,yshift=.75cm,align=center,text depth=.75cm,font=\small},
}

\titleformat{\section}{\large\bfseries}{\thesection.}{0.5em}{}
\titleformat{\subsection}{\normalsize\bfseries}{\thesubsection.}{0.5em}{}

\theoremstyle{plain}
\newtheorem{theorem}{Theorem}[section]
\newtheorem{proposition}[theorem]{Proposition}
\newtheorem{lemma}[theorem]{Lemma}

\theoremstyle{definition}

\theoremstyle{remark}

\usepackage[textsize=tiny]{todonotes}

\newcommand{\R}{ \mathbb{R}}

\newcommand{\norm}[1]{\left\rVert#1\right\rVert}

\def\V#1{{\boldsymbol{#1}}}         
\def\M#1{{\bf{#1}}}  

\renewcommand{\[}{\begin{equation}}
\renewcommand{\]}[1]{\label{eq:#1}\end{equation}}

\DeclareMathOperator*{\argmin}{arg\,min}

\def\BibTeX{{\rm B\kern-.05em{\sc i\kern-.025em b}\kern-.08em T\kern-.1667em\lower.7ex\hbox{E}\kern-.125emX}}

\newcommand{\euler}{\mathrm{e}}

\title{\Large DEALing with Image Reconstruction: Deep Attentive Least Squares}
\author{Mehrsa Pourya$^{\dagger}$, Erich Kobler$^{\ddagger}$, Michael Unser$^{\dagger}$, Sebastian Neumayer$^{\star}$ 
\thanks{\texttt{mehrsa.pourya@epfl.ch}, \texttt{erich.kobler@jku.at}, 
 \\ \texttt{michael.unser@epfl.ch}, \texttt{sebastian.neumayer@\\mathematik.tu-chemnitz.de}}
\\ \small $\dagger$ Biomedical Imaging Group, EPFL Lausanne, Switzerland 
\\ \small $\ddagger$ Institute for Machine Learning, LIT AI lab, Institute for Virtual Morphology, Johannes Kepler University Linz
\\ \small $\star$ Faculty of Mathematics, TU Chemnitz, Germany
}

\date{}

\begin{document}

\maketitle

\begin{abstract}
State-of-the-art image reconstruction often relies on complex, highly parameterized deep architectures.
We propose an alternative: a data-driven reconstruction method inspired by the classic Tikhonov regularization.
Our approach iteratively refines intermediate reconstructions by solving a sequence of quadratic problems.
These updates have two key components: (i) learned filters to extract salient image features, and (ii) an attention mechanism that locally adjusts the penalty of filter responses.
Our method achieves performance on par with leading plug-and-play and learned regularizer approaches while offering interpretability, robustness, and convergent behavior.
In effect, we bridge traditional regularization and deep learning with a principled reconstruction approach.
\end{abstract}

\section{Introduction}
Image reconstruction plays a fundamental role in computational imaging and computer vision \cite{MM2019, zeng2001image}.
The task is to recover an unknown image of interest $\M x \in \R^d$ from noisy measurements $\M y \in \R^M$.
Their relation is often modeled as $\M y = \M H \M x$, where the forward operator $\M H \in \R^{M \times d}$ encodes the acquisition process. 
If $\M H$ is ill-conditioned, one resorts to regularized reconstruction
\begin{equation} \label{eq:general_inverse_problem}
    \hat {\M x} \in \argmin_{\M x \in \R^d} \norm{\M H \M x - \M y}_2^2 + \lambda \mathcal{R}(\M x).
\end{equation}
The data-fidelity $\norm{\M H \M x - \M y}_2^2$ controls the consistency of the reconstruction with the measurements.
The regularizer $\mathcal{R}\colon \R^d \to \R_{\geq 0}$ encodes prior information about the solution and is also intended to make the problem well-posed.
Both terms are balanced by the hyperparameter $\lambda \in \R_{\geq 0}$.
Throughout this paper, $\M x$ is the vectorized version of a (color or grayscale) image with shape $N_{\mathrm{in}}\times H \times W$.

From classic signal processing to the advent of deep learning, a significant body of research focuses on the design of the regularizer $\mathcal{R}$. 
In the context of data-driven methods, two primary approaches emerged: (i) the explicit modeling of $\mathcal{R}$, and (ii) the modeling of operators associated with $\mathcal{R}$, such as its proximal operator, which is required in Plug-and-Play (PnP) reconstruction algorithms \citep{venkatakrishnan2013plug}.
Following the explicit approach, the starting point for this work is the 
fields-of-experts model \citep{RotBla2009}, which reads
\begin{equation}\label{eq:rridge_reg_mask}
	\mathcal{R}_{\M m} \colon\M{x}\mapsto \sum_{c=1}^{ N_\mathrm{C}} \bigl \langle 
    \M m_c, \psi_c (\M W_{c} \M x)\bigr\rangle.
\end{equation}
Here, each $\M W_c \in \R^{HW \times d}$ convolves $\M x$ with a filter template $w_c \in \R^{N_{\mathrm{in}} \times k_s \times k_s}$.
Then, the nonnegative potentials $\psi_c \in \mathcal{C}(\R)$ are applied entry-wise to the $\M W_c \M x$.
Finally, the weights $\M m_c \in [\epsilon_M, 1]^{HW}$ with $\epsilon_M>0$ determine the (spatially varying) contribution of $\psi(\M W_c \M x)$ to the regularizer.
In principle, every component of \eqref{eq:rridge_reg_mask} can be learned.
Most implementations so far have used constant weights $\M m_c = \M 1$ in \eqref{eq:rridge_reg_mask}.
Recent works differ in the parameterization of $\psi_c$ and $\M W_c$ for the learning process \citep{CheRan2014, GouNeuUns2023, ZaKo23ProductGMDM} or incorporate non-linear feature transforms \citep{LiSch2020, koblerTotalDeepVariation2022}.
While learning $\M W_c$ and $\psi_c$ is extensively studied, little research is performed regarding the local weights $\M m_c$.

Employing spatially varying $\M m_c$ in \eqref{eq:rridge_reg_mask} is referred to as anisotropic regularization.
For instance, $\M m_c = \M M(\M y)$ can be derived from the data $\M y$ using heuristics \citep{chanInpaintingFlexibleHaarWavelet2008,grasmairAnisotropicTotalVariation2010} or a neural network \citep{KofAltBa2023,lefkimmiatis2023learning}.
When $\M M$ extracts features from an estimated reconstruction, as proposed by \citet{NeuAlt2024}, it is natural to consider refining $\M m_c$ iteratively.
Specifically, the reconstruction from~\eqref{eq:general_inverse_problem} can be fed back into $\M M$ to obtain an \emph{improved} $\M m_c$ for \eqref{eq:rridge_reg_mask}. 
This leads to the attentive reconstruction process
\begin{equation} \label{eq:iterative_refinement}
    \M x_{k+1} \in \argmin_{\M x \in \R^d} \norm{\M H \M x - \M y}_2^2 + \lambda \mathcal{R}_{\M M(\M x_k)}(\M x),
\end{equation}
studied by \citet{PouNeuUns2024} for $\psi_c=|\cdot|$.
We note that their scheme is computationally demanding since each update involves a least absolute shrinkage and selection operator (LASSO)-type problem.

\paragraph{Contribution}
The updates \eqref{eq:iterative_refinement} are the starting point of our investigation.
Our contributions are as follows.
\begin{itemize}
     \item We simplify the updates \eqref{eq:iterative_refinement} by choosing $\psi_c(\cdot) = (\cdot)^2$ for the $\mathcal R_{\M M(\M x_k)}$ from \eqref{eq:rridge_reg_mask}.
    Then, each update amounts to solving a linear equation, which we handle efficiently through the conjugate gradient method.
    \item We learn the components of $\mathcal{R}_{\M M(\M x_k)}(\M x)$, namely the filters \smash{$\{\M W_c\}_{c=1}^{N_c}$} and the attention mechanism $\M M$, based on a denoising task in such a way that the iterations \eqref{eq:iterative_refinement} converge.
    Then, any fixed point of \eqref{eq:iterative_refinement} is a consistent reconstruction. 
    Given a linear inverse problem with forward $\M H$, we then apply our learned model with only two remaining hyperparameters: (i) the noise level and (ii) the regularization strength $\lambda$.
    \item On the theoretical side, we establish the uniqueness of each update in \eqref{eq:iterative_refinement}, the existence of a fixed point, a condition for the convergence of \eqref{eq:iterative_refinement}, and a stability result for the resulting reconstruction operator.
    \item In our experimental evaluation, we achieve results on par with state-of-the-art approaches for various inverse problems. 
    \item We underline the interpretability, convergence, and robustness of our method. We provide illustrative examples to visualize the learned attention mechanism. 
\end{itemize}

\section{Related Literature}

\paragraph{Learned Regularization}
Classical regularizers for \eqref{eq:general_inverse_problem} leverage sparsity in various domains, such as image gradients \citep{rudin1992nonlinear} or wavelets \citep{mallatWaveletTourSignal1999}.
The parametric model \eqref{eq:rridge_reg_mask} introduced by \citet{RotBla2009} has since spurred extensive research.
Key areas of investigation include learning paradigms \citep{CheRan2014, EffKob2020}, parametrization strategies \citep{ZaKo23ProductGMDM}, and intrinsic properties like convexity \citep{GouNeuBoh2022, GouNeuUns2023}.
More complex architectures build upon strategies such as autoencoders \citep{LiSch2020}, algorithm unrolling \citep{koblerTotalDeepVariation2022}, adversarial training \citep{lunz2018adversarial, prostLearningLocalRegularization2021}, and energy modeling \citep{zachStableDeepMRI2023}.

In parallel, implicit regularization methods were developed.
Here, of-the-shelf denoisers are incorporated into iterative reconstruction algorithms by, for instance, replacing proximal operators \citep{venkatakrishnan2013plug, Drunet2022} or drawing inspiration from adaptive Laplacians \citep{romano2017little}.
If the denoiser is non-expansive or homogeneous, which is hard to ensure for learned ones \citep{reehorst2018regularization,HNS2021}, this leads to a convergent scheme.
Recently, weaker conditions have been proposed by \citet{pesquetLearningMaximallyMonotone2021a, hurault2022gradient}.

\paragraph{Spatial Adaptivity}
An overview of spatially adaptive regularizers with the form \eqref{eq:rridge_reg_mask} is given in \citet{PraCalLan2023}.
The authors \citet{HiPaRa2017,ChReSc2017,KofAltBa2023} focus on the total variation regularizer \citep{rudin1992nonlinear} as a special case, using either heuristics or deep learning to compute the weights $\M m_c$.
More general instances are considered by \citet{lefkimmiatis2023learning,NeuPouGou2023,NeuAlt2024}.
The first work deploys non-smooth potentials $\psi_c$ and majorization minimization to solve the nonsmooth problem \eqref{eq:general_inverse_problem}.
Similar to \eqref{eq:iterative_refinement}, this leads to a series of quadratic problems.
The latter works deploy differentiable $\psi_c$ and solve \eqref{eq:general_inverse_problem} with accelerated gradient descent.
All approaches have in common that they update the weights $\M m_c$ only once.
In particular, they do not refine $\M m_c$ and the reconstruction iteratively as in \eqref{eq:iterative_refinement}.

\paragraph{Iterative Refinement}
\citet{LenLelBec2014,LenBer2015} propose to iteratively refine the weights $\M m_c$ for total variation.
They update the $\M m_c$ using some heuristic.
For the more general model \eqref{eq:rridge_reg_mask} with $\psi_c = \vert \cdot \vert$, a refinement based on neural networks was considered by \citet{PouNeuUns2024}.
Their $\M M$ has a simple architecture comparable to ours.
Outside of this setting, iterative refinement can be found, for example, in \citet{SahHoCha2023} for superresolution and in \citet{DarNatLi2024} for MRI.

\paragraph{Nonlocal Laplacians}
Quadratic potentials $\psi_c$ lead to an optimality condition for \eqref{eq:iterative_refinement} induced by a symmetric positive semi-definite matrix.
A different approach to getting such updates is (iterative) filtering with carefully designed graph Laplacians \citep{PanChe2017}.
Recently, this idea was incorporated into deep architectures for image denoising \citep{ZenPanSun2019,ValFraMag2020} and scene flow \cite{TeDe21}.


\section{Methodology}
We specify the regularizer \eqref{eq:rridge_reg_mask} using quadratic potentials as 
\begin{equation}
    \label{eq:als_reg_simple}
    \mathcal{R}_{\M m}(\M x) = \sum_{c=1}^{N_c} \bigl\langle \M m_c^2, (\M W_c \M x)^2\bigr\rangle = \norm{\M M \M W \M x}_2^2
\end{equation}
with the shorthands $\M W=[\M W_1^\top \ \ldots \ \M W_{N_c}^\top]^\top$ and $\M M = \mathrm{Diag}(\mathrm{Diag} (\M m_1),\ \ldots,\ \mathrm{Diag} (\M m_{N_c}))$, where $\mathrm{Diag}$ returns a diagonal matrix whose entries are the input vector.
This leads to the quadratic reconstruction problem
\begin{equation} \label{eq:tikh_like}
   \hat{\M x} \in \argmin_{\M x \in \R^d} \norm{\M H \M x - \M y}_2^2 + \lambda \norm{\M M \M W \M x}_2^2.
\end{equation}
Now, the question arises about the proper choice of $\M M$  and $\M W$.
Regarding a filter-based interpretation, the $\M W$ should be independent of $\V y$.
$\M M$ modulates the response $\M W \M x$ at each location.
Ideally, this modulation should depend on the structure of the (unknown) $\hat{\M x}$.
With abuse of notation, we introduce $\M M\colon \R^d \to [\epsilon_M,1]^{N_cHW}$, which leads to our attentive refinement scheme
\begin{align}\label{eq:IterRefine}
    \M x_{k+1} &\in \mathcal{T}(\M x_k, \M y) \quad \text{with} \quad  \M x_0 = \M 0 \in \R^d,\\
    \mathcal{T}(\M z, \M y) & = \argmin_{\M x \in \R^d}  \norm{\M H \M x - \M y}_2^2 + \lambda \norm{\M M(\M z) \M W \M x}_2^2.\label{eq:tikh_iterative}
\end{align}
The process \eqref{eq:IterRefine} can be interpreted as an infinite-depth neural network.
Hence, we name it deep attentive least squares (DEAL) for image reconstruction.
If $\M x_{k} \to \hat{\M x}$, we get $\hat{\M x} \in \mathcal{T}(\hat{\M x}, \M y)$, namely that $\hat{\M x}$ is a fixed-point of the operator \eqref{eq:tikh_iterative}.
We restrict $k < K_{\mathrm{out}}$ and terminate the iterations \eqref{eq:IterRefine} when $\norm{\M x_{k+1} - \M x_{k}}_2/\norm{\M x_{k}}_2 \leq \epsilon_{\mathrm{out}}$ with $\epsilon_{\mathrm{out}} > 0$.

\subsection{Architecture}
\begin{figure}[tbp]
\begin{center}
    \includegraphics[width=8cm,height=5cm,keepaspectratio]{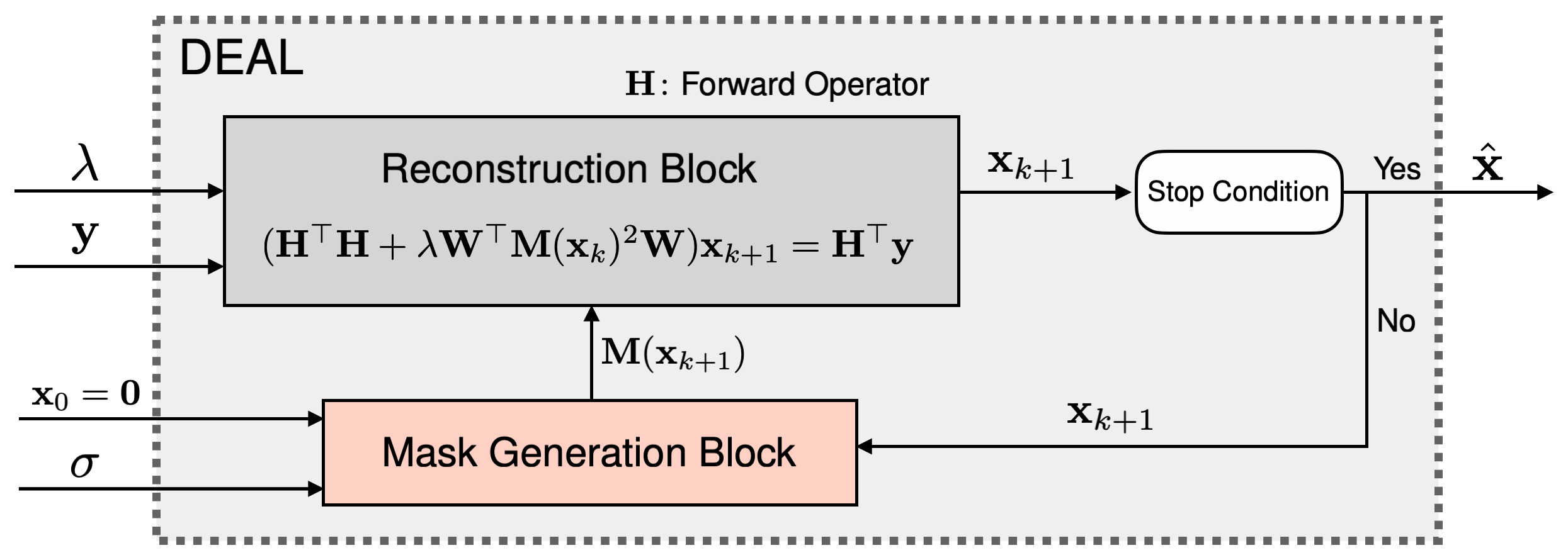}
  \caption{DEAL generates a sequence of reconstructions $\M x_k$ via \eqref{eq:IterRefine} from the inputs $\M y$ and $\M H$, initalization $\M x_{0} = \M 0$, and hyper-parameters $\sigma$ and $\lambda$.
  When the stop condition is met, it returns $\hat{\M x}$.}
  \label{fig:main_arch}
\end{center}
\end{figure}

Next, we specify how the DEAL iterates~\eqref{eq:IterRefine} can be cast as a deep neural network structure.
Figure \ref{fig:main_arch} visualizes the interplay of its essential building blocks -- reconstruction and mask generation -- which exchange information repeatedly.

\subsubsection{Reconstruction Block}
At the heart of DEAL, the reconstruction block solves the spatially-adapted optimization problem \eqref{eq:tikh_iterative} for given attentive weights $\M M(\M x_k)$ (see Section \ref{sec:MaskBlock}).
The optimality condition for problem \eqref{eq:tikh_iterative} is given by the linear system
\begin{equation}\label{eq:UpdateEquation}
    \M A_{k} \M x_{k+1} = \M b
\end{equation}
with $\M A_{k} = \M H^{\top} \M H +  \lambda \M W ^{\top} \M M(\M x_k)^2 \M W$ and $\M b = \M H^{\top} \M y$.
The multi-convolution block $\M W$, see Section~\ref{sec:ConvBlock},  and $\lambda \in \R$ are learnable.
To avoid scaling ambiguities, we impose $\norm{\M W}_2 = 1$ by spectral normalization.
The data $\M y$ and the forward $\M H$ are problem-specific inputs that are not learnable.
We solve \eqref{eq:UpdateEquation} by the conjugate-gradient (CG) algorithm with $\M x_k$ as the initial guess.
We use a batched CG with at most $K_{\mathrm{in}}$ steps, where each sample terminates individually if its residue satisfies $\norm{\M A_k - \M b \M x_{k+1}}_2^2 \leq \epsilon_{\mathrm{in}}$ for $\epsilon_{\mathrm{in}} > 0$.

\subsubsection{Mask Generation Block}\label{sec:MaskBlock}
\begin{figure}[tbp]
\begin{center}
  \includegraphics[width=8cm,height=5cm,keepaspectratio]{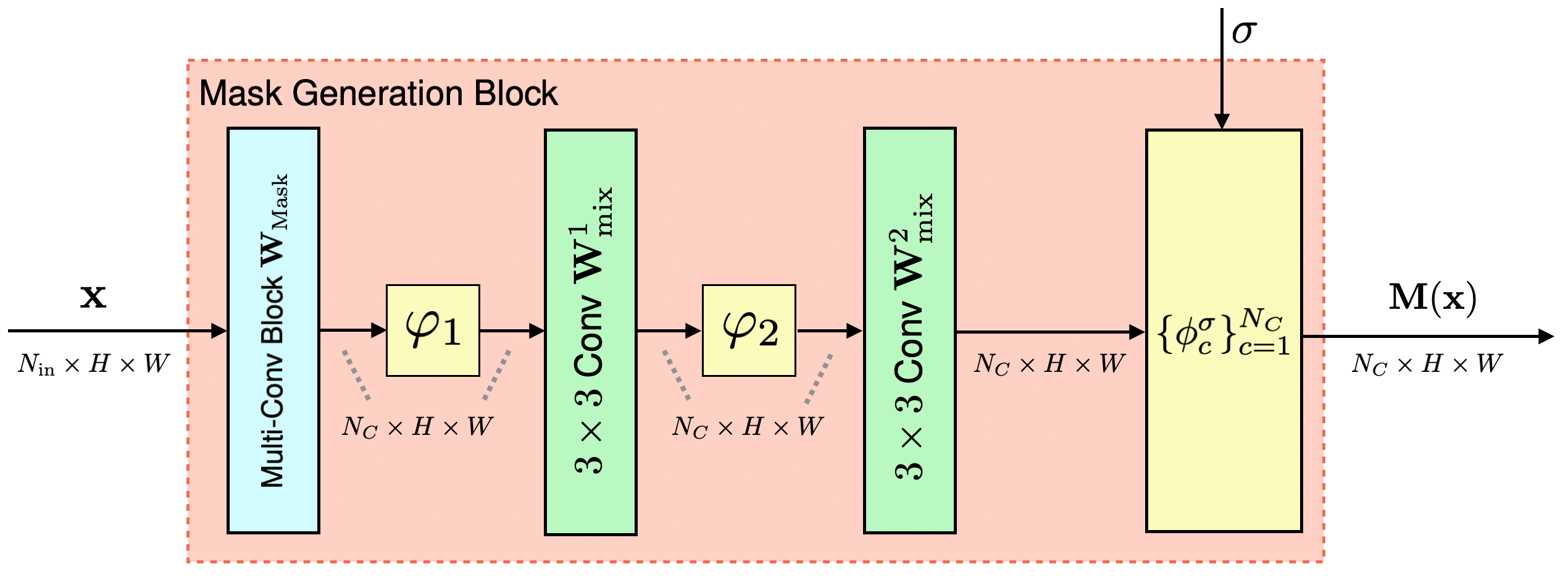}
  \caption{Architecture of the mask generation block.}
  \label{fig:mask_arch}
\end{center}
\end{figure}

To estimate $\M M$ for \eqref{eq:tikh_like} from local image structures, we use the following shallow CNN with learnable nonlinearities
\begin{equation} \label{eq:MaskNet}
    \M M(\M x) = (\boldsymbol \phi^{\sigma} \circ \M W_\mathrm{mix}^2 \circ \varphi_2 \circ \M W_\mathrm{mix}^1 \circ \varphi_1\circ\M W_\mathrm{mask})(\M x),
\end{equation}
see also Figure \ref{fig:mask_arch}.
This choice is inspired by anisotropic diffusion \citep{weickertAnisotropicDiffusionImage1998,brediesMathematicalImageProcessing2018}, where $\M M$ is typically computed from the gradients of a smoothed image using pixel-wise non-linearities.

The first multi-conv layer $\M W_\mathrm{mask}$ (see Section \ref{sec:ConvBlock}) in \eqref{eq:MaskNet} extracts $N_c$ spatial features using the same architecture as the $\M W$ from the reconstruction block.
The two subsequent convolution layers $\M W_\mathrm{mix}^1$ and $\M W_\mathrm{mix}^2$ mix the $N_c$ feature channels using kernels of size $3\times3$ and no bias.
The layers are connected via learnable point-wise non-linearities $\varphi_1$ and $\varphi_2$, for which we follow \citet{BCGA2020}.
Specifically, each $\varphi_i$ is parametrized as a linear spline with $N_n$ equally distributed knots on $[0, r]$.
On $(r,\infty)$ the splines are linearly extended and we enforce symmetry by setting $\varphi(x) = \varphi(-x)$ if $x < 0$.
In addition, we constrain them to be increasing for $x > 0$.
Removing both constraints has not led to significant performance improvements.

To guarantee the numerical stability of \eqref{eq:UpdateEquation}, the outputs of $\M M$ must remain in $(\epsilon_{M},1]$.
Hence, we process each channel $c\in\{1\ldots N_c\}$ individually by\begin{equation}
    \phi^{\sigma}_c(x) = \max\left(\min\left(\varphi_3(\alpha_c(\sigma)\M x) ,1 \right), \epsilon_{M}\right),
\end{equation}
where $\varphi_3$ is a symmetric linear spline as before. 
In contrast to the former splines, $\varphi_3$ must be non-increasing on $[0,\infty)$.
The underlying rationale is that the $\M M(\M x)$ should be close to $1$ for small filter responses (constant image regions) and close to $0$ for strong filter responses (salient edges).
Following \citet{GouNeuUns2023}, we enable multi-level noise training using the positive scalings
\begin{equation}
    \label{eq:scalings_}
    \alpha_{c}(\sigma) = \frac{\euler^{s_{c}(\sigma)}}{\sigma + 10^{-5}}
\end{equation}
with learnable linear splines $\{s_{c}\}_{c=1}^{N_C}$. 
By design of $\M M$, the first reconstruction block consists of a non-varying problem.

\subsubsection{Multi-Conv Block}\label{sec:ConvBlock}
The Multi-Conv block advocated by \citet{GouNeuBoh2022} consists of multiple convolution layers with no non-linearities in between.
It enables the efficient construction of large receptive fields.
We use three convolution layers, all with kernels of size $(9 \times 9)$.
Thus, the effective field of view for this block is $(25 \times 25)$.
For inputs with $N_{\mathrm{in}}$ channels, the number of output channels of the convolution layers are  $4 N_{\mathrm{in}}$, $8 N_{\mathrm{in}}$, and $N_{C}$, respectively.
We set the group size and stride to one, and do not use a bias.
In all our experiments, we use $N_{C} = 128$. 
This block appears in two places: once in $\M M(\M x)$ as $\M W_{\mathrm{Mask}}$ and second in the reconstruction block as $\M W$, i.e., the convolutions in \eqref{eq:als_reg_simple}.

\subsection{Training}
We learn the parameters $\theta$ of DEAL for image denoising with additive white Gaussian noise (AWGN) of varying standard variation $\sigma_n \in [0,50]$.
The associated denoiser $D_{\theta(\sigma_n)}^{K_\mathrm{out}}(\M y)$ takes the measurements $\M y$ and $\sigma_n$ as input, and returns the solution of \eqref{eq:IterRefine} with at most $K_{\mathrm{out}}$ iterations. 
Choosing denoising as the training task has two reasons: (i) our learned model should also work for other inverse problems (often called universality), as demonstrated by \citet{hurault2022gradient,GouNeuUns2023}; (ii) it simplifies the computations for the reconstruction block as $\M H = \M I$. We provide details on initializations and hyperparameters with a short ablation study in \ref{app_hypp}. 

\paragraph{Dataset and Loss}
For the training dataset $\mathsf{D}=\{\M x^m\}_{m=1}^M$, we use the images proposed in \citet{Drunet2022}.
The images $\M x$ are corrupted by AWGN as $\M y = \M x + \sigma_n \M n$ and fed into DEAL, leading to a sequence of denoised images $\{\M x_k\}_{k=1}^{K_\mathrm{out}} = D_{\theta(\sigma_n)}^{K_\mathrm{out}}(\M y)$.
To estimate the parameters $\theta$ from the training data, we use the loss
\begin{align} \label{eq:loss}
    \mathcal{L}(\theta) &= \Bigg\{ \mathbb{E}_{\substack{\M x\sim\mathsf{D}\\\sigma_n \sim\mathcal{U}([0, 50])\\\M n \sim\mathcal{N}(\M 0, \M I)}}\Big[ \norm{\M x_{K_\mathrm{out}} - \M x}_2^2 +\\ \nonumber &\frac{\gamma}{N_c}\norm{\M M(\M x_{K_\mathrm{out}})-\M M(\M x_{K_\mathrm{out}-1})}_2^2 \Big] + \gamma\mathrm{TV}^2(\theta) \Bigg\}
\end{align}
with $\gamma=10^{-4}$.
This loss consists of three parts: (i) a squared error enforcing that the output matches the clean image; (ii) a squared penalty on the weight changes for the last two updates of \eqref{eq:IterRefine}; and (iii) an accumulated second-order total variation regularization of all learnable splines.
This last penalizes changes in their slopes (kinks) and thereby promotes simpler splines \cite{BCGA2020, DGB2022}.
The second term in \eqref{eq:loss} vanishes if the generated weights $\M M(\M x_k)$ converge.
To promote convergence of \eqref{eq:IterRefine} to a fixed point, we sample $K_{\mathrm{out}}$ uniformly from $[15, 60]$ \citep{AniAshKai2022}.

We minimize the loss \eqref{eq:loss} using Adam \citep{kingma2014adam}.
At each step of the optimizer, we sample $16$ patches of size $(128\times 128)$ randomly from $\mathsf{D}$.
We have two training phases; first, we train the gray and color model for $\num{70000}$ and $\num{40000}$ steps respectively, using an initial learning rate $\num{5e-4}$ that is reduced to $\num{4e-4}$ by a cosine annealing scheduler.
Then, we continue training the gray and color model for $\num{10000}$ and $\num{5000}$ steps, respectively, with an initial learning rate $\num{2e-4}$ that is reduced to $\num{1e-7}$ by annealing.
We set $\epsilon_{\mathrm{out}} = \epsilon_{\mathrm{in}} = \num{1e-4}$ and the maximal of CG steps to $K_{\mathrm{in}} = 50$.
To select the best-performing model, we evaluate its performance every $\num{1000}$ training step and keep the checkpoint with the best validation performance.
We use the set3 and set12 datasets to validate the color and grayscale models.

\paragraph{Gradient Tracking}
We train DEAL through the deep equilibrium framework \citep{BKK2019} in the Jacobian free mode.
Specifically, we perform at most  $K_{\mathrm{out}}-1$ iterations without gradient tracking.
Then, after convergence, we perform one additional update \eqref{eq:IterRefine} with gradient tracking.
For this, it is crucial to have an efficient backward path for the reconstruction block, namely for $\partial_\theta \M x_{k+1}$ as defined in \eqref{eq:IterRefine}.
Since backpropagating through the CG algorithm is prohibitively memory extensive, we need another solution.

From \eqref{eq:UpdateEquation}, we have that
\begin{equation}
   \M x_{k+1} = \M y -  \lambda \M L_k^{\top} \M L_k \M x_{k+1},
\end{equation}
where $\M L_k = \M M(\M x_k) \M W$. 
Using the product rule, we get 
\begin{equation}
    \partial_\theta \M x_{k+1} = -  \lambda \M L_k^{\top} \M L_k \M \partial_\theta \M x_{k+1} - \lambda \partial_\theta (\M L_k^{\top} \M L_k) \M x_{k+1}.
\end{equation}
It follows
\begin{equation}
\label{eq:backward_linear_system}
   \M A_k \partial_\theta \M x_{k+1} = \M d_{k+1},
\end{equation}
where the matrix $\M A_k$ is the one from \eqref{eq:UpdateEquation}, and only the right-hand-side changes to $\M d_{k+1} = - \lambda \partial_\theta ( \M L_k^{\top} \M L_k) \M x_{k+1}$.
We use auto-differentiation to obtain the gradient estimate $\M d_{k+1}$. We then find $\partial_\theta \M x_{k+1}$ by solving \eqref{eq:backward_linear_system} with CG.

         
\subsection{Inference}
Once the parameters are learned, we can directly deploy DEAL to a general inverse problem by plugging the corresponding forward operator $\M H$ and its adjoint $\M H^{\top}$ into the reconstruction block.
This does not affect the mask generation block.
To adapt to the new task, we now only need to tune two hyperparameters: the model noise level $\sigma$ and the regularization strength $\lambda$ in \eqref{eq:IterRefine}.
This requires a small validation set with paired measurements and ground truth images.
\begin{figure*}[htb]
    \centering
    \resizebox{\linewidth}{!}{
    \begin{tikzpicture}
    \foreach \f/\l/\m [count = \xi from 0] in {
        img_clean/Original/,
        measurements/Measurement/,
        img_wcrr/WCRR/(29.51 0.85),
        img_safi/SAFI/(29.99 0.86),
        img_deal/DEAL (Ours)/(30.07 0.86),
        img_prox_drunet/ProxDRUNet/(30.13 0.85),
        img_drunet/DRUNet/(30.49 0.88)
        }
    {
        \begin{scope}[spy using outlines={rectangle, draw=white, magnification=3.5, width=2.4cm, height=1.75cm}]
        \node[img,label={[mylabel]\l \\ \m}] (i\xi) at (2.5*\xi,0) {\includegraphics[width=2.4cm]{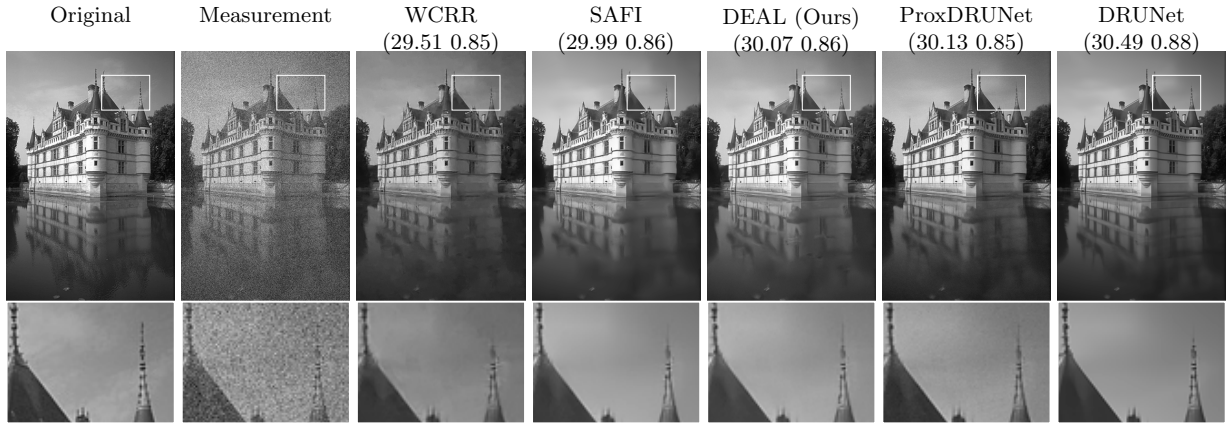}};
        \spy on (2.5*\xi+.5,1.2) in node (z) [below] at (i\xi.south);
        \end{scope}
    }
    \end{tikzpicture}}
    \vspace{-.2cm}
    \caption{Denoising of the \emph{castle} image for $\sigma = 25$. For each reconstruction (PSNR and SSIM) is provided. }
    \label{fig:castle_compare}
\end{figure*}
Empirically, we observe that tuning $\lambda$ is more important than changing the noise level $\sigma$.
To ensure that we find a fixed point of \eqref{eq:IterRefine}, we set $K_{\mathrm{in}} = K_{\mathrm{out}} = 1000$, and choose the conservative stop criteria $\epsilon_{\mathrm{in}} =10^{-8}$ and $\epsilon_{\mathrm{out}} = 10^{-5}$. 

\section{Theoretical Results}\label{sec:Theory}
All the proofs are provided in Appendix~\ref{sec:proofs}.
Proposition~\ref{prop:UniqueSol} guarantees the uniqueness of the updates \eqref{eq:IterRefine}.
\begin{proposition}\label{prop:UniqueSol}
    If $\ker (\M H) \cap \ker (\M M(\M x_k) \M W) = \{\M 0\}$, then $\M A_k$ is positive definite and \eqref{eq:UpdateEquation} has a unique solution.
    Moreover, if $\M M^2(\M x_k) \succeq \epsilon_M \mathrm{Id}$, then $\M A_k \succeq \M H^\top \M H + \epsilon_M \M W^\top \M W $ and uniqueness holds if $\ker (\M H) \cap \ker ( \M W) = \{\M 0\}$.
\end{proposition}
The next results involve an estimate of the smallest eigenvalue of $\M A_k \colon$  $\lambda_\epsilon = \lambda_\mathrm{min}(\M H^\top \M H + \epsilon_M \M W^\top \M W)$. 
\begin{lemma}\label{lem:LipData}
    Let $\M x \in \R^d$.
    If $\ker (\M H) \cap \ker (\M W) = \{\M 0\}$ and $\M M(\M x)^2 \succeq \epsilon_M \mathrm{Id}$, then $\mathcal{T}(\M x, \cdot) \colon \R^M \to \R^d$ is Lipschitz continuous with constant $\Vert \M H \Vert_2/ \lambda_\epsilon$.
\end{lemma}
Next, we establish the existence of fixed points for $\mathcal T(\cdot, \M y)$.
\begin{theorem}\label{thm:ExistFix}
    Assume $\ker (\M H) \cap \ker (\M W) = \{\M 0\}$ and $\M M^2 (\M x)\succeq \epsilon_M \mathrm{Id}$.
    Then, $\mathcal T(\cdot, \M y)\colon \R^d \to \ B_r(\M 0) $ maps into a ball around $\V 0$ with radius $r = \norm{\M H \M y }_2/\lambda_\epsilon$.
    If $\M M^2 \colon B_r(\M 0) \to [\epsilon,1]^{N_\mathrm{out} HW}$ is Lipschitz continuous with constant $L$, then $\mathcal T(\cdot, \M y)$ admits a fixed point and
    \begin{equation}\label{eq:LipEst}
    	\Vert \mathcal{T}(\M x_1, \M y) - \mathcal{T}(\M x_2, \M y)\Vert_2  \leq \frac{L \norm{\M H \M y }_2}{\lambda^2_\epsilon} \Vert \M x_2 - \M x_1 \Vert_2.
    \end{equation}
\end{theorem}
The Lipschitz estimate \eqref{eq:LipEst} is very conservative and $\mathcal T(\cdot, \M y)$ appears to often be even a local contraction.
If $\mathcal T(\cdot, \M y)$ is contractive for every $\M y \in \R^M$, we get the following result.
\begin{theorem} \label{thm:last}
    Assume $\ker (\M H) \cap \ker (\M W) = \{\M 0\}$ and $\M M^2 (\M x)\succeq \epsilon_M \mathrm{Id}$.
    If $\mathcal{T}(\cdot, \M y) \colon \R^d \to \R^d$ is contractive, i.e., if $\Vert \mathcal{T}(\M x_1, \M y) - \mathcal{T}(\M x_2, \M y) \Vert_2 \leq q \Vert \M x_1 - \M x_2\Vert_2$ with $q < 1$, then the iterations \eqref{eq:IterRefine} converge to a unique fixed point $\hat {\M x}$ and
    \begin{equation}\label{eq:ExpConv}
        \Vert \M x_k - \hat {\M x}\Vert_2 \leq q^{k-1} \Vert \M x_1 - \M x_0 \Vert_2.
    \end{equation}
    In particular, we have exponential convergence of \eqref{eq:IterRefine}.
    Moreover, if $\hat {\M x} = \mathcal{T}(\hat {\M x}, \M y_1)$ and $\hat {\M z} = \mathcal{T}(\hat {\M z}, \M y_2)$, then it holds
    \begin{equation}\label{eq:StabEstimate}
        \Vert \hat {\M x} - \hat {\M z} \Vert \leq \frac{1}{1-q}\frac{\Vert \M H \Vert_2}{\lambda_\epsilon} \Vert \M y_1 - \M y_2 \Vert_2.
    \end{equation}
\end{theorem}

\section{Experiments}
We compare our method to state-of-the-art PnP approaches and learned regularizers on denoising, super-resolution, and MRI reconstruction.
In Appendix \ref{app:debluring}, we provide experimental results for grayscale debluring.  

\subsection{Grayscale and Color Denoising}
\begin{figure*}
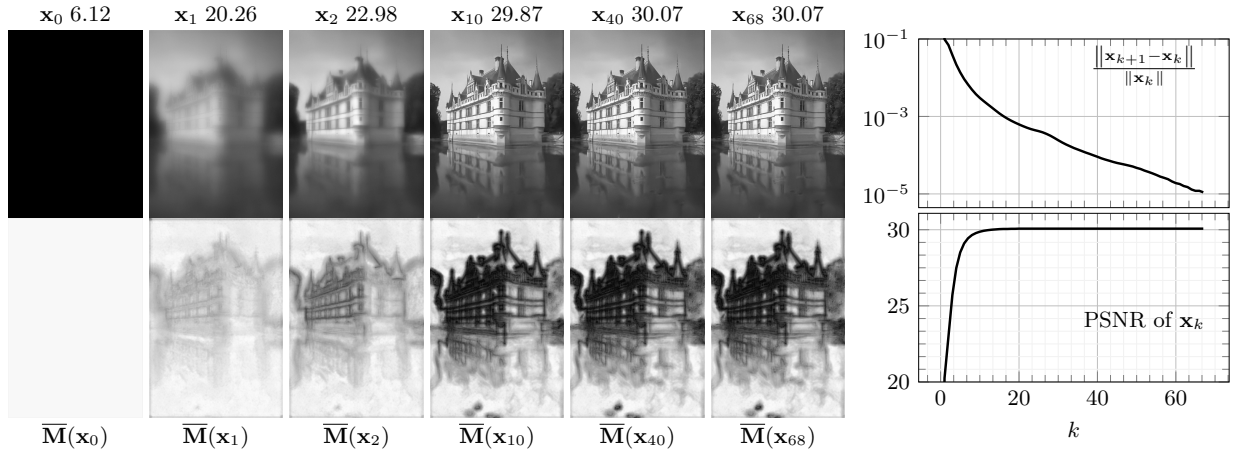

    \centering
    \resizebox{\linewidth}{!}{
    \begin{tikzpicture}
    \foreach \i/\m [count = \xi from 0] in {
        0/6.12,
        1/20.26,
        2/22.98,
        10/29.87,
        40/30.07,
        68/30.07
        }
    {
        \node[img,label={[mylabel,yshift=-.3cm]$\M x_{\i} \; \m$}] (i\xi) at (2*\xi,0) {\includegraphics[width=1.9cm]{Files_for_erich/fig_4_castle_path/sol_\i.png}};
        \node[img,label={[mylabel,yshift=-.7cm]below:$\overline{\M M}(\M x_{\i})$}] (i\xi) at (2*\xi,-2.7) {\includegraphics[width=1.9cm]{Files_for_erich/fig_4_castle_path/mask_\i.png}};
    }

    \begin{scope}[xshift=12cm,yshift=-1.2cm,font=\small]
        \begin{axis}[%
                anchor=north west,
                table/col sep=comma,
                width=6cm,
                height=4cm,
                title=PSNR of $\M x_k$,
                title style=
                {
                  xshift=1cm,
                  yshift=-2cm,
                },
                xlabel={$k$},
                ylabel={},
                grid=both,
                grid style={line width=.1pt, draw=gray!10},
                major grid style={line width=.2pt,draw=gray!50},
                minor tick num=5,
                ymin=20,
            ]
            \addplot[line width=1pt,mark=none] table[x=x,y=psnr] {Files_for_erich/fig_4_castle_path/psnr.csv};
        \end{axis}
    \end{scope}
    \begin{scope}[xshift=12cm,yshift=1.3cm,font=\small]
        \begin{semilogyaxis}[%
                anchor=north west,
                table/col sep=comma,
                width=6cm,
                height=4cm,
                title=$\frac{\norm{\M x_{k+1} -\M x_k}}{\norm{\M x_k}}$,
                title style=
                {
                  xshift=1cm,
                  yshift=-1cm,
                },
                ylabel={},
                ymax=0.1,
                grid=both,
                grid style={line width=.1pt, draw=gray!10},
                major grid style={line width=.2pt,draw=gray!50},
                minor tick num=5,
                xmajorticks=false,
            ]
            \addplot[line width=1pt,mark=none] table[x=x,y=res] {Files_for_erich/fig_4_castle_path/psnr.csv};
        \end{semilogyaxis}
    \end{scope}
    \end{tikzpicture}}
    \vspace{-1.0cm}
    \caption{Solution path and channel-wise averages $\overline{\M M}$ of the weights for DEAL iterations, exemplified with the \emph{castle} image and $\sigma_n = 25$.
    On the right, we plot the residual values and PSNR over the number of outer iteration $k$.}
    \label{fig:castle_path}
\end{figure*}
We corrupt ground-truth images by adding white Gaussian noise with standard deviation $\sigma_n \in \{5, 15, 25\}$. 
\begin{table}[htb]
        \centering
        \small
        \caption{Denoising for the BSD68 and CBSD68 datasets.}%
        \vspace{.1cm}
     \label{table:denoising_gray}
        \setlength\tabcolsep{2pt}
        \begin{tabular}{lcccccc}
        \toprule
        & \multicolumn{3}{c}{Gray} & \multicolumn{3}{c}{Color} \\
        $\sigma_n$ & 5 & 15 & 25 & 5 & 15 & 25 \\
        \toprule
        BM3D & 37.54 & 31.13 & 28.61 & 39.89 & 33.17 & 30.38\\
        \midrule
        WCRR & 37.65 & 31.20& 28.68 & $-$ & $-$ &$-$ \\
        SARR &  37.80 & 31.61 & 29.13 & $-$ & $-$ &$-$ \\
        SAFI & 37.90 & 31.56 & 29.05 & $-$ & $-$ &$-$\\
        DEAL (Ours) &  37.85 & 31.61 &  29.16  & 40.04 & \underline{33.61} & \underline{30.93}  \\
        Prox-DRUNet &  \underline{37.97}&  31.70& 29.18 & \underline{40.12} &  33.60 & 30.82\\

        \midrule
        DNCNN &  $-$ &\underline{31.72}&  \underline{29.23} & 39.80 & 33.55 & 30.87\\
        DRUNet &  \textbf{38.09} &  \textbf{31.94}& \textbf{29.48} & \textbf{40.19} & \textbf{33.85} & \textbf{31.21}\\
        \bottomrule
        \end{tabular}

        \centering
         \small
        \vspace{0.5cm}
        
        \caption{PSNR values for color-image superresolution.}%
        \label{table:superres}
        \vspace{.1cm}
        \setlength\tabcolsep{2pt}
        \begin{tabular}{lcccccccc}
        \toprule
        & \multicolumn3{c}{s = 2} & \multicolumn{3}{c}{s=3} \\
        & 2.55 & 7.65 & 12.75 & 2.55 & 7.65 & 12.75 \\
        \toprule
         DEAL (Ours) & \textbf{27.99} &   \underline{26.58}&   25.75& \textbf{26.20}& \underline{25.27}  &24.59  \\
          Prox-DRUNet  & 27.88 & \textbf{26.61} & \underline{25.79} & \underline{26.13} & \textbf{25.29} & \textbf{24.67}\\
        \midrule
        IRCNN  & 26.97 & 25.86 & 25.45 & 25.60 & 25.72 & 24.38\\
        DPIR   & 27.79 & \underline{26.58} & \textbf{25.83} & 26.05 &  \underline{25.27} & \underline{24.66} \\
        \bottomrule
\end{tabular}
\end{table}
For various methods, we provide the respective average peak signal-to-noise ratios (PSNR) over the images of the BSD68 set and the center-cropped (256x256) images of CBSD68 set in Table \ref{table:denoising_gray}.
Since some approaches are implemented for grayscale images only, parts of the table are left blank.
First, we include (C)BM3D \cite{DabFoiKat2007} as a widely regarded classic baseline.
We also evaluate against WCRR \cite{GouNeuUns2023}, an unadaptive field-of-expert model of the form \eqref{eq:rridge_reg_mask} that employs weakly convex potentials $\psi_c$.
We also include its data-adaptive counterpart SARR \cite{NeuPouGou2023,NeuAlt2024}.
Regarding the refinement perspective \eqref{eq:iterative_refinement}, we include SAFI \cite{PouNeuUns2024}, which utilizes $\psi_c = \ell_1$ instead of $\ell_2$.
Lastly, we include the deep learning-based approaches DnCNN \cite{ZhaZuoChe2017}, DRUNet \cite{Drunet2022} and Prox-DRUNet \cite{HurLec2022}.
The latter trades performance for improved theoretical guarantees compared to DRUNet. 
DEAL outperforms existing spatially adaptive methods and closes the gap to DRUNet-based approaches while having 30 times fewer parameters. 
We provide qualitative results in Figure~\ref{fig:castle_compare}, where we also provide the structural similarity index metric (SSIM).
In the magnified part, we can see that DEAL does better than the DRUNet-based approaches in retaining structures such as the \textit{tip} of the tower.
In Figure~\ref{fig:castle_path}, we provide the solution path associated to \eqref{eq:IterRefine}, the averages of the masks $\M M(\M x_k)$, and two convergence plots.
Specifically, the weights $\M M$ extract the image structure, leading to lower regularization cost at edges.
\begin{table}[tbp]
        \centering
        \small
        \caption{PSNR values for the MRI experiment.}%
        \vspace{.1cm}
        \label{table:reconstruction_performance_mri}
        \setlength\tabcolsep{2pt}
        \begin{tabular}{lcccc}
        \toprule
        & \multicolumn{2}{c}{4-fold single coil} & \multicolumn{2}{c}{8-fold multi-coil} \\
        & PD & PDFS & PD & PDFS \\
        \midrule
        Zero-fill ($\M H^{\top} \V y$) & 27.40 & 29.68  & 23.80 & 27.19  \\
        TV & 32.44 & 32.67 & 32.77 & 33.38  \\
        WCRR  & 35.78 & 34.63 &  35.57 & 35.16 \\
        SARR & 36.25 & 34.77 & 35.98 & 35.26  \\
        SAFI  & \underline{36.43}& \underline{34.92} & \underline{36.06}  & \textbf{35.36} \\
        DEAL (Ours)  &  \textbf{36.59}&  \underline{34.92}&  \textbf{36.21}&  \underline{35.32}\\
        Prox-DRUNet  & 36.20 & \textbf{35.05} & 35.82 & 35.12 \\
         PnP-DnCNN  & 35.24 & 34.63 &  35.11 & 35.14 \\
        \bottomrule 
        \end{tabular}
\end{table}

\subsection{Color Super-resolution}
\begin{figure}[htbp]
    \centering
    \resizebox{0.94\linewidth}{!}{
    \begin{tikzpicture}
    \foreach \f/\l/\m [count = \xi from 0] in {
        img_clean/Original/,
        measurements/Measurement/,
        img_deal/DEAL (Ours)/(28.21 0.83)/ 
        }
    {
        \begin{scope}[spy using outlines={rectangle, draw=white, magnification=4, width=2.6cm, height=1.65cm}]
        \node[img,label={[mylabel, yshift=0.1cm]\l \\ \m}] (i\xi) at (2.7*\xi,0) {\includegraphics[width=2.6cm]{Files_for_erich/fig_5_superres_comparison/\f.png}};
        \spy on (2.7*\xi-.6,-.6) in node (z) [below] at (i\xi.south);
        \end{scope}
    }
    \node[img,anchor=north west] at (i0.north west) {\includegraphics[width=1cm]{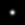}};
    \foreach \f/\l/\m [count = \xi from 0] in {
        img_prox_drunet/ProxDRUNet/(27.95 0.82),
        img_ircnn/IRCNN/(27.27 0.79),
        img_dpir/DPIR/(27.99 0.83)
        }
    {
        \begin{scope}[spy using outlines={rectangle, draw=white, magnification=4, width=2.6cm, height=1.65cm}]
        \node[img,label={[mylabel,yshift=-.75cm]below:\l \\ \m}] (i\xi) at (2.7*\xi,-6) {\includegraphics[width=2.6cm]{Files_for_erich/fig_5_superres_comparison/\f.png}};
        \spy on (2.7*\xi-.6,-.59-6) in node (z) [above] at (i\xi.north);
        \end{scope}
    }
    \end{tikzpicture}}
    \vspace{-.45cm}
    
    \centering
    \resizebox{0.94\linewidth}{!}{
    \begin{tikzpicture}
    \begin{scope}[font=\small]
        \begin{axis}[%
                table/col sep=comma,
                width=6cm,
                height=8cm,
                ymax=32,
                title=PSNR of $\M x_k$,
                xlabel={$k$},
                label style={yshift=.2cm},
                ylabel={},
                restrict y to domain=0:40,
                grid=both,
                grid style={line width=.1pt, draw=gray!10},
                major grid style={line width=.2pt,draw=gray!50},
                minor tick num=5,
                cycle list name=exotic,
                legend entries={$\M x_0=\M 0$\\$\M x_0\sim\mathcal{U}([0,1])$\\$\M x_0\sim \mathcal{N}(0,\M I)$\\$\M x_0=\M H^\top \M y$ \\ $\M x_0=\M x$\\},
                legend pos=south east,
                legend cell align={left},
            ]
            \addplot[color=blue!50,line width=1pt,mark=*] table[x=k,y=zero] {Files_for_erich/fig_6_superres_robust/psnr_curves.csv};
            \coordinate (zero) at (axis cs:0,7.15) {};
            \addplot[color=red,line width=1pt,mark=x] table[x=k,y=uniform] {Files_for_erich/fig_6_superres_robust/psnr_curves.csv};
            \coordinate (uniform) at (axis cs:0,9.23) {};
            \addplot[color=green!75!black,line width=1pt,mark=star] table[x=k,y=normal] {Files_for_erich/fig_6_superres_robust/psnr_curves.csv};
            \coordinate (normal) at (axis cs:0,7.24) {};
            \addplot[color=purple,line width=1pt,mark=triangle] table[x=k,y=adm] {Files_for_erich/fig_6_superres_robust/psnr_curves.csv};
            \coordinate (adm) at (axis cs:0,9.89) {};
            \addplot[color=cyan!75,line width=1pt,mark=diamond] table[x=k,y=gt] {Files_for_erich/fig_6_superres_robust/psnr_curves.csv};
            \coordinate (gt) at (axis cs:0,32) {};
            \coordinate (converged) at (axis cs:28,28.21);
        \end{axis}
    \end{scope}
    \foreach \f/\c [count = \xi from 0] in {zero/blue!50,normal/green!75!black,uniform/red,adm/purple,gt/cyan!75}
    {
        \node[img,draw=\c,ultra thick] (i\f) at (-2,.4+\xi*1.35) {\includegraphics[width=1.25cm]{Files_for_erich/fig_6_superres_robust/\f_00.png}\includegraphics[width=1.25cm]{Files_for_erich/fig_6_superres_robust/\f_mask_00.png}};
        \draw[\c,ultra thick] (i\f.east) -- (\f);
    }
    \node[xshift=-.6cm,minimum height=1.75em,anchor=north] at (izero.south) {$\M x_0$};
    \node[xshift=.7cm,minimum height=1.75em,anchor=north] at (izero.south) {$\overline{\M M}(\M x_0)$};
    \node[img,draw=black,ultra thick] (iconverged) at (3,4) {\includegraphics[width=1.25cm]{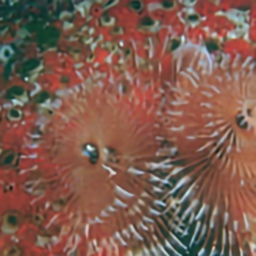}\includegraphics[width=1.25cm]{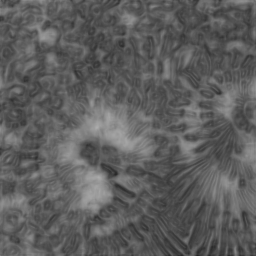}};
    \draw[black,ultra thick] (iconverged.north) -- (converged);
    \end{tikzpicture}}
    \vspace{-0.1cm}
    \caption{Superresolution task with rate $s=2$ and $\sigma_n=2.55$.
    The bottom image illustrates the dependence on initialization.}
    \label{fig:supperres_compare}
    \vspace{.1cm}

     \centering
     \small
    \captionof{table}{Computation time (seconds) for the MRI experiment.}%
    \label{table:mri_time}
    \vspace{.1cm}
        \setlength\tabcolsep{2pt}
        \begin{tabular}{lcccc}
        \toprule
        & \multicolumn{2}{c}{4-fold single coil} & \multicolumn{2}{c}{8-fold multi-coil} \\
        & PD & PDFS & PD & PDFS \\
        \midrule
        WCRR  &  \textbf{12} & \textbf{20} & \textbf{9} & \textbf{8}\\
        SAFI  &436 &  470& 388& 326\\
        DEAL (Ours) & \underline{14} &  \underline{17}& \underline{22}& \underline{18}\\
        Prox-DRUNet  & 113 & 38 & 170 & 105\\

        \bottomrule
        \end{tabular}
\end{figure}

Here, the forward $\M H$ involves two steps: the bluring of the image through the convolution with a known kernel; followed by a downsampling with $s \in \{2, 3\}$ that reduces the number of measurements by a factor of $s^2$.
As further degradation, AGWN of standard deviation $\sigma_n$ is added to the image.
We deploy the four Gaussian blur kernels from \cite{Drunet2022} with standard deviations ($0.7$, $1.2$, $1.6$ and $2.0$), and report the average over these for the center-cropped images of CBSD68 in Table \ref{table:superres}.
There, DPIR \cite{Drunet2022} represents the state-of-the-art PnP approach; and IRCNN \cite{zhang2017learning} is a competing PnP approach. 
The hyperprameters of DEAL are set to $\sigma=15$ and $\lambda \in \{0.28, 2.5, 5.5\}$ for the three given noise levels $\sigma_n \in \{2.55, 7.65, 12.75\}$, respectively.
We outperform existing methods on the lower noise levels and achieve comparable performance for the other cases.
In Figure \ref{fig:supperres_compare}, we provide a visual comparison.

\begin{figure}[tbp]
\centering
\includegraphics[width=\linewidth]{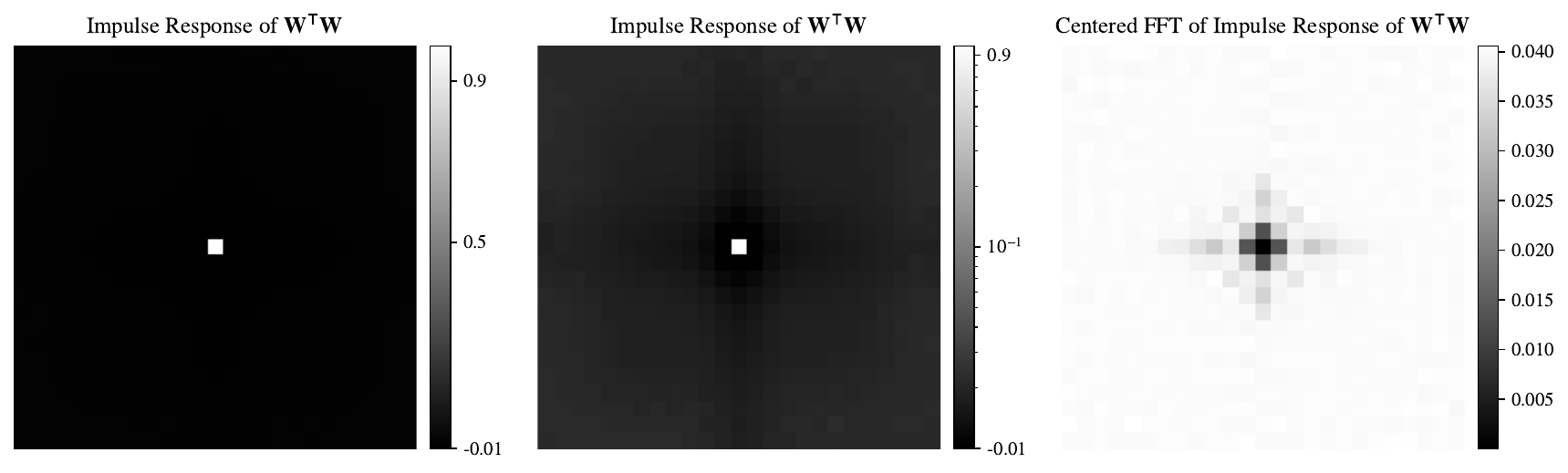}
\caption{Different visualizations of $\M W^\top \M W$.
Only one eigenvalue is numerically zero. The middle plot is on a logarithmic scale.}
\label{fig:kernels}
\end{figure}
 
\subsection{MRI Reconstruction}
Now, we deploy DEAL for magnetic resonance imaging (MRI) tasks.
Specifically, we tackle the single- and 15-coil MRI setups detailed by \citet{GouNeuBoh2022}.
There, the ground-truth consists of knee images from the fastMRI dataset \cite{zbontar2018fastMRI}, both with fat suppression (PDFS) and without fat suppression (PD).
The forward $\M H$ involves $k$-fold subsampling in the Fourier domain and corruption by additive white Gaussian noise with $\sigma_n = 0.002$.
For each of the four evaluation tasks, we use ten images to tune the hyperparameters of all methods.
In Table~\ref{table:reconstruction_performance_mri}, we report the PSNR values on centered $(320\times 320)$ patches of the remaining fifty test images.
We compare against the popular TV regularization, the convex CRR regularizer \cite{GouNeuBoh2022}, its weakly convex extension WCRR, and the Prox-DRUNet.
All methods are \emph{universal} in that they can be deployed without task-specific training.
In Table \ref{table:mri_time}, we report the computation times for several methods on a Tesla V100-SXM2-32GB GPU.
We are significantly faster than the iterative refinement approach SAFI and get close to the (non-adaptive) WCRR baseline.
Qualitative results are given in Figure \ref{fig:compare_mri} of Appendix \ref{app:vis_mri}.

\section{Interpretability and Robustness}
In Figures~\ref{fig:supperres_compare} and \ref{fig:mri_path}, we demonstrate empirically that DEAL is not tied to a specific number $K_\mathrm{out}$ of iterations \eqref{eq:IterRefine}.
In particular, doing more updates does not degrade the performance, unlike many PnP methods such as DPIR.
The convergence to a fixed point (see Theorem \ref{thm:ExistFix}) occurs for all experiments.
In particular, both the relative error and the PSNR converge.  
To demonstrate the robustness regarding initialization, we instantiate the superresolution task. As we see in Figure~\ref{fig:supperres_compare}, DEAL converges (in about 10 steps) to the same reconstruction independent of the initialization. This is in accordance with Theorem \ref{thm:last}. 
The relative errors for this task are given in Figure \ref{fig:convergence_multires} of Appendix \ref{app_conv_superees}, underlining once more the empirical convergence.

\begin{figure}
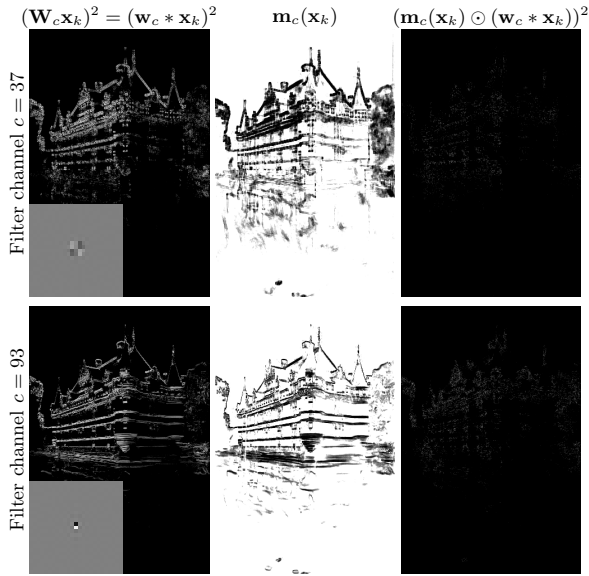

    \centering
    \resizebox{\linewidth}{!}{
    \begin{tikzpicture}
    \foreach \k [count = \yi from 0] in {93,37}
    {
        \foreach \f [count = \xi from 0] in {Wx2,mask,Lx2}
        {
            \node[img] (n\xi) at (3*\xi,4.5*\yi) {\includegraphics[width=2.9cm]{Files_for_erich/last_figure_mask_stuff/filter_\k_\f.png}};
        }
        \node[img, anchor=south west] at (n0.south west) {\includegraphics[width=1.5cm]{Files_for_erich/last_figure_mask_stuff/filter_\k_f.png}};
        \node[rotate=90,yshift=2mm,font=\small] at (n0.west) {Filter channel $c=\k$};
    }
    \node[yshift=2mm,font=\small] at (n0.north) {$(\M W_c \M x_k)^2 = (\M w_c\ast \M x_k)^2$};
    \node[yshift=2mm,font=\small] at (n1.north) {$\M m_c(\M x_k)$};
    \node[yshift=2mm,font=\small] at (n2.north) {$(\M m_c(\M x_k) \odot (\M w_c \ast \M x_k))^2$};
    \end{tikzpicture}}
    \caption{From left to right: two example filters $\M W_c$ along with $(\M W_c \M x_k)^2$; the corresponding masks $\M m_c(\M x_k)$; and adapted squared response.
    The mask eliminates the structure of the image in the squared responses.
    Hence, DEAL preserves salient structures.}
    \label{fig:mask_per_channel}
\end{figure}

We present visualizations for all parts of our architecture in Appendix \ref{app_under_the_hood}. 
Remarkably, we find mostly finite differences and their higher-order counterparts at various scales within $\M W$ (see Figure \ref{fig:W_gray_filters} in the appendix).
These filters extract the salient features of the input.
The impulse response of $\M W^{\top} \M W$ and its Fourier transform are given in Figure \ref{fig:kernels}.
Empirically, we observe that $\ker (\M W) = \mathrm{span}(\mathbf 1_d)$.
This is a practical certification for the condition of Proposition \ref{prop:UniqueSol}. 

Next, we present two interpretations of the learned attention mechanism.
For simplicity, we focus on the castle denoising example from Figure \ref{fig:castle_compare} with the final solution $\M x_k$.
In Figure \ref{fig:mask_per_channel}, we illustrate two learned filters within $\M W_c$ and the response $\M W_c \M x_k$ to these filters.
The associated weights (masks) $\M m_c(\M x_k)$ are well adapted to the structural features captured by these filters.
In effect, the mask suppresses the image structures in the final squared responses $(\M m_c(\M x_k) \odot (\M w_c \ast \M x_k))^2$, leading to a reduced regularization cost in \eqref{eq:als_reg_simple}.
This is desirable as the image structure should not contribute to the regularization cost.

\begin{figure}[tbp]
    \centering
    \resizebox{\linewidth}{!}{
    \begin{tikzpicture}
    \node at (3*0,1.75) {$\M y$};
    \node at (3*1,1.75) {$\M x_{k} = \M A_{k-1}^{-1} \M y$};
    \node at (3*2,1.75) {$\M A_{k-1}^{-1} \M e_n$};
    \foreach \f [count = \xi from 0] in {y,gt,ak}
    {
        \foreach \y in {0,1,2}
        {
            \node[img] at (3*\xi,-3*\y) {\includegraphics[width=2.9cm]{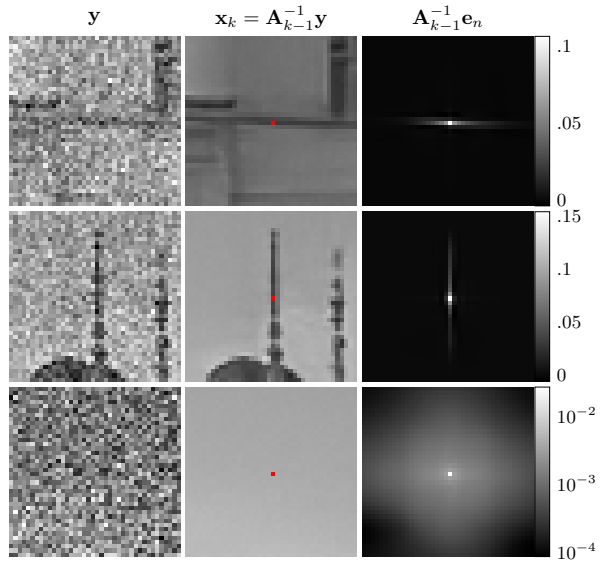}};
        }
    }
    \begin{scope}[xshift=7.3cm]
    \begin{axis}[
        hide axis,
        scale only axis,
        height=0pt,
        width=0pt,
        colormap/blackwhite,
        colorbar horizontal,
        point meta min=0,
        point meta max=0.11,
        colorbar style={
            width=2.9cm,
            height=0.25cm,
            rotate=90,
            xtick={0,0.05,0.1},
            xtick style={draw=none},
            xticklabels={0,.05,.1},
            xticklabel style={xshift=0ex,yshift=1mm,anchor=west,font=\small},
            anchor=center,
        }
        ]
        \addplot [draw=none] coordinates {(0,0) (1,1)};
    \end{axis}
    \end{scope}
    \begin{scope}[xshift=7.3cm,yshift=-3cm]
    \begin{axis}[
        hide axis,
        scale only axis,
        height=0pt,
        width=0pt,
        colormap/blackwhite,
        colorbar horizontal,
        point meta min=0,
        point meta max=0.16,
        colorbar style={
            width=2.9cm,
            height=0.25cm,
            rotate=90,
            xtick={0,0.05,0.1,0.15},
            xtick style={draw=none},
            xticklabels={0,.05,.1,.15},
            xticklabel style={xshift=0ex,yshift=1mm,anchor=west,font=\small},
            anchor=center,
        }
        ]
        \addplot [draw=none] coordinates {(0,0) (1,1)};
    \end{axis}
    \end{scope}
    \begin{scope}[xshift=7.3cm,yshift=-6cm]
    \begin{axis}[
        hide axis,
        scale only axis,
        height=0pt,
        width=0pt,
        colormap/blackwhite,
        colorbar horizontal,
        point meta min=0.,
        point meta max=1,
        colorbar style={
            width=2.9cm,
            height=0.25cm,
            rotate=90,
            xtick={0,.4,0.8},
            xtick style={draw=none},
            xticklabels={$10^{-4}$,$10^{-3}$,$10^{-2}$},
            xticklabel style={xshift=0ex,yshift=1mm,anchor=west,font=\small},
            anchor=center,
        }
        ]
        \addplot [draw=none] coordinates {(0,0) (1,1)};
    \end{axis}
    \end{scope}
    \end{tikzpicture}}
    \caption{The reconstruction $\M x_{k}$ (middle) at the $n$-th entry (in red) is the weighted average of the data $\M y$ (left) with the weights (right) that come from the matrix $\M A_{k-1}^{-1}$, i.e., $\M x_{k}[n] = \langle \M A_{k-1}^{-1} \M e_n, \M y\rangle$.
    Top to bottom: three different regions of the castle image.}
    \label{fig:graph}
\end{figure}
Next, we describe our second interpretation.
Since we are in the denoising setting ($\M H = \M I$), we have $\M x_{k} = \M A_{k-1}^{-1}  \M y$
with $\M A_{k-1} = \M I +  \lambda \M W ^{\top} \M M(\M x_{k-1})^2 \M W$.
Therefore, the mapping from the measurements $\M y$ to the solution $\M x_k$ is simply a linear transformation.
More precisely, the $n$-th component of $\M x_k$ is a weighted average of the measurements $\M y$ with the weight given by the $n$-th row of $\M A^{-1}_{k-1}$. 
To extract this row, we apply \smash{$\M A^{-1}_{k-1}$} to the $n$-th unit vector $\M e_n$.
Since $\M x$ and $\M y$ are vectorized images, we reshape the rows accordingly for Figure \ref{fig:graph}. 
There, we see that $\M A_{k-1}^{-1} \M e_n$ aligns well with the structure of the neighborhood around the $n$-th pixel of the image.
This indicates that spatial information is encoded into $\M A_{k-1}$ during the refinements.
In the first and second rows, the averaging occurs in the vertical and horizontal directions, respectively.
In contrast, the image of the third row exhibits little structure, and DEAL averages over a larger region, with an emphasis on the center pixel.
Thus, at equilibrium, DEAL acts as an adaptive averaging mechanism, intelligently averaging the noisy measurements $\M y$, with weights that emerge from our iterative refinements.


\section{Conclusion}
We have presented deep attentive least squares (DEAL) for image reconstruction.
DEAL builds upon classic signal processing ideas, which we blended with recent advances in deep learning, particularly, infinite-depth networks.
It consists of two parts: (i) an iterative refinement of intermediate reconstructions based on a least-square-type problem; and (ii) a recurrent attention mechanism that adapts the problem spatially.
We achieved competitive performance on different tasks while being able to provide interpretability, universality, and theoretical guarantees.

So far, we only trained DEAL on the denoising task.
If sufficient data is available, it appears possible to fine tune all components of DEAL to further improve its performance.
Moreover, DEAL is designed for the $\ell_2$ data fidelity, and devising extensions for other data-fidelity terms is an interesting direction of future work. 




\section*{Acknowledgements}
M.P. and M.U. acknowledge support from the European Research Council (ERC Project FunLearn) under Grant 101020573, in part by the Swiss National Science Foundation, Grant $200020\_219356$.
S.N.\ and E.K.\ acknowledge support from the DFG within the SPP2298 under project number 543939932.
E.K acknowledges support from the Austrian Science Fund (FWF) project number 10.55776/COE12.

\bibliography{dhtv,references,references_cvx,refs}
\bibliographystyle{plainnat}

\newpage
\appendix
\onecolumn

\section{DEAL Hyperparameters \label{app_hypp} and Ablation Studies}
For the linear splines $\varphi_1$, $\varphi_2$ and $\varphi_3$ appearing in $\M M$ we fix $r = 3$, $N_n = 31$.
Moreover, we initialize $\varphi_1$ and $\varphi_2$ as the absolute value $\vert \cdot \vert$ and $\varphi_3$ as $\euler^{-(\cdot)^2}$, motivated by classical anisotropic diffusion \citep{weickertAnisotropicDiffusionImage1998}.
Each $s_c$ in \eqref{eq:scalings_} has $14$ knots in the range $[-1, 51]$.
They are initialized as the constant function $3$.
We choose $\sigma = \sigma_n$ in $\M M$, where $\sigma_n$ is the standard deviation of a sample's noise.
We set $\lambda = \kappa(\sigma_n)$ with a learnable spline $\kappa$ defined using  52 knots in $[-1, 51]$ initialized as the identity.
The denoisers are strongly tied, particularly, $\M W$ and $\M M$ must work across various settings. 

We used $N_c = 128$ number of filters. We observed that if we reduce $N_c$ to 64 or 32, we degrade the performance for denoising on $\sigma = 25$ by around $0.1$ and $0.15$, respectively.
We also observed that going beyond the filter size $(9 \times 9)$ in the Multi-Conv block does not improve the performance.
The proposed constraints and initializations for the learning of the non-linearites stem from the learning of such parts without any constraints and with zero initialization. 

\section{Proofs}\label{sec:proofs}
\subsection{Proof of Proposition \ref{prop:UniqueSol}}
\begin{proof}
    Assume that there exists $\M x \in \R^d \setminus \{\M 0\} \in \ker(\M A_k)$, namely with $\M x^\top \M A_k \M x = \M 0$.
    By definition of $\M A_k$, this implies $\M x^\top \M H^\top \M H \M x = \M 0$ and $\M x^\top \M W^\top \M M^2 (\M x_k) \M W \M x = \M 0$.
    Hence, we get $\M x \in \ker (\M H)$ and $\M x \in \ker (\M M (\M x_k)\M W)$, which is a contradiction.
    For $\M M^2(\M x_k) \succeq \epsilon_M \mathrm{Id}$, we estimate $\M A_k \succeq \M H^\top \M H + \epsilon_M \M W^\top \M W $ and the uniqueness as in the first part.
\end{proof}
\subsection{Proof of Lemma \ref{lem:LipData}}
\begin{proof}
    Let $\hat {\M x} = \mathcal{T}(\M x, \M y_1)$ and $\hat {\M z} = \mathcal{T}(\M x, \M y_2)$, namely $\M A_{\M x} \hat {\M x} = \M H^\top \M y_1$ and $\M A_{\M x} \hat {\M z} = \M H^\top \M y_2$.
    This implies that $\M A_{\M x} (\hat {\M x} - \hat {\M z}) = \M H^\top (\M y_1 - \M y_2)$ and we estimate
    \begin{equation}
        \Vert \hat {\M x} - \hat {\M z} \Vert_2 \leq \frac{\Vert \M H \Vert_2}{\lambda_\epsilon} \Vert \M y_1 - \M y_2 \Vert_2. \qedhere
    \end{equation}
\end{proof}
\subsection{Proof of Theorem \ref{thm:ExistFix}}
\begin{proof}
    First, we investigate the range of $\mathcal T(\cdot, \M y)$.
    By definition of $\mathcal T(\cdot, \M y)$, it holds for any $\M x \in \R^d$ that 
    \begin{equation}\label{eq:RangeImage}
        \Vert \mathcal T(\M x, \M y) \Vert_2 = \Vert \M A_k^{-1} \M H^\top \M y \Vert_2 \leq \frac{\norm{\M H \M y }_2}{\lambda_\epsilon}.
    \end{equation}
    For the second part, we want to apply Brouwer's fixed-point theorem.
    To this end, we must prove that $\mathcal T(\cdot, \M y)$ is continuous.
    Let $\M x_1, \M x_2 \in \R^d$, $\hat {\M x}_1 = \mathcal{T}(\M x_1, \M y)$ and $\hat {\M x}_2 = \mathcal{T}(\M x_2, \M y)$.
    Then, it holds
    \begin{align}
        \M A_1 \hat{\M x}_1 - \M A_2 \hat{\M x}_2 &= \M 0 \notag\\
        \M A_1 \hat{\M x}_1 - \M A_1 \hat{\M x}_2 &= \M A_2 \hat{\M x}_2 - \M A_1 \hat{\M x}_2 \notag\\
         \hat{\M x}_1 - \hat{\M x}_2 &= \M A_1^{-1} (\M A_2 - \M A_1) \hat{\M x}_2.
    \end{align}
    Incorporating \eqref{eq:RangeImage} and the normalization $\Vert \M W \Vert_2= 1$, we further infer
    \begin{align}
         \Vert \hat{\M x}_1 - \hat{\M x}_2 \Vert_2
         \leq &\Vert \M A_1^{-1} \Vert_2 \Vert \M W \Vert_2^2 \Vert\M M^2(\M x_2) - \M M^2 (\M x_1) \Vert_2 \Vert \hat{\M x}_2 \Vert_2\notag\\
          \leq&  \frac{L \norm{\M H \M y }_2}{\lambda^2_\epsilon} \Vert \M x_2 - \M x_1 \Vert_2.
    \end{align}
    Hence, $\mathcal T(\cdot, \M y)$ is Lipschitz continuous and a fixed-point exists.
\end{proof}
\subsection{Proof of Theorem \ref{thm:last}}
\begin{proof}
	Due to the Banach fixed point theorem, the exponential convergence rate \eqref{eq:ExpConv} holds. 
	To estimate the difference of $\hat {\M x} = \mathcal{T}(\hat {\M x}, \M y_1)$ and $\hat {\M z} = \mathcal{T}(\hat {\M z}, \M y_2)$, we use the contractivity of $\mathcal{T}(\cdot, \M y_1)$ and Lemma \ref{lem:LipData} to get
	\begin{align}
		\Vert \hat {\M x} - \hat {\M y} \Vert_2 = \Vert \mathcal{T}(\hat {\M x}, \M y_1) - \mathcal{T}(\hat {\M z}, \M y_2)\Vert_2
		\leq & \Vert \mathcal{T}(\hat {\M x}, \M y_1) - \mathcal{T}(\hat {\M z}, \M y_1)\Vert_2 + \Vert \mathcal{T}(\hat {\M z}, \M y_1) - \mathcal{T}(\hat {\M z}, \M y_2)\Vert_2\notag\\
		\leq & q \Vert \hat {\M x} - \hat{\M z} \Vert_2 + \frac{\Vert \M H \Vert_2}{\lambda_\epsilon}  \Vert \M y_1 - \M y_2 \Vert_2.
	\end{align}
	From this, we readily infer \eqref{eq:StabEstimate}, namely
	\begin{equation}
		\Vert \hat {\M x} - \hat {\M y} \Vert_2 \leq \frac{\Vert \M H \Vert_2}{(1-q)\lambda_\epsilon} \Vert \M y_1 - \M y_2 \Vert_2. \qedhere
	\end{equation}
\end{proof}

\section{Additional Numerical Results}

\subsection{Grayscale Deblurring \label{app:debluring}}

\begin{figure*}
    \centering
    \includegraphics[width=\linewidth]{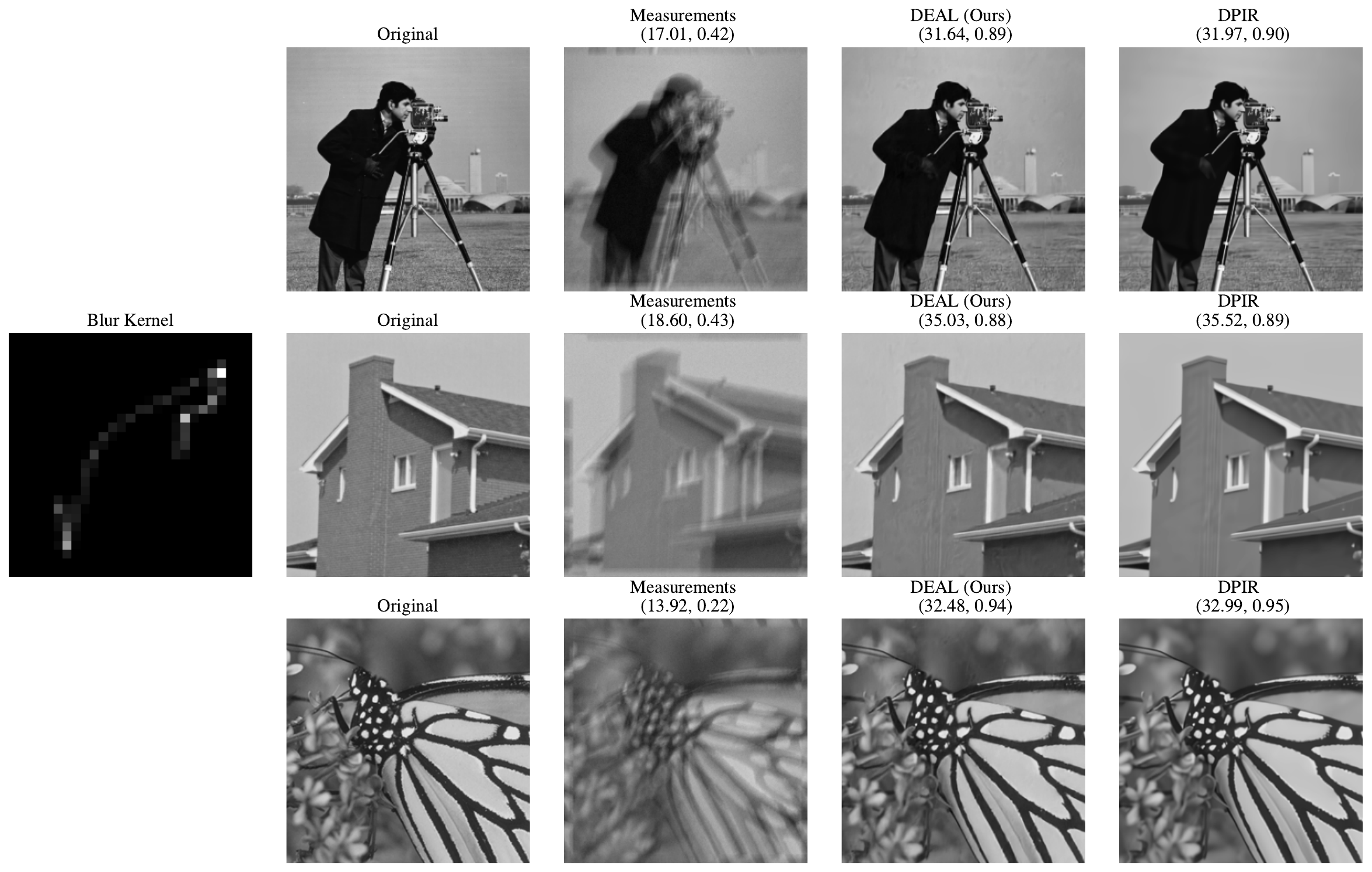}
    \caption{Debluring results for the given blur kernel and noise level $\sigma_n = 2.55$. }
    \label{fig:debluring}
\end{figure*}

Here, we evaluate the DEAL approach for a grayscale deblurring task.
We use the same setup as DPIR for this experiment \cite{Drunet2022}.  This includes two blur kernels of sizes $17 \times 17$ and $27 \times 27$ from \cite{levin2009understanding} and additive Gaussian noise with $\sigma_n=2.55$ and $\sigma_n=7.65$.
In Table \ref{table:delur}, we report the PSNR of the reconstructions for the Set3 images, namely Cameraman, House, and Monarch.
We set the model noise level $\sigma=15$ and $\lambda \in \{0.5, 2.5\}$ for the two given AWGN noise levels $\sigma_n$, respectively.
We also compare with model-based EPLL \cite{zoran2011learning} and the learning approach FDN that is specific to debluring \cite{kruse2017learning}. We observe that we are consistently the second-best method on this task after DPIR. We provide a visual comparison with DPIR in Figure \ref{fig:debluring}. 
\subsection{Visualizations for MRI Reconstruction \label{app:vis_mri}}
\begin{figure*}[htbp]
    \centering
    \includegraphics[width=1\linewidth]{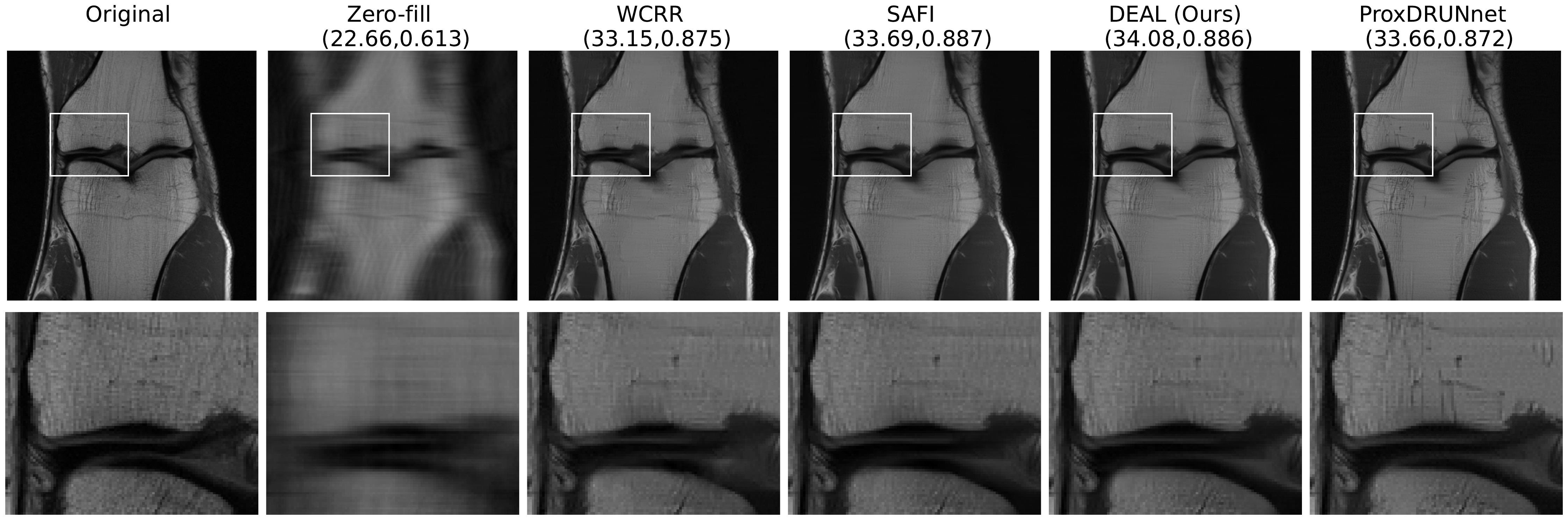}
    \caption{Comparison of different methods for 8-fold multi-coil MRI reconstruction for a PD image of the knee.}
    \label{fig:compare_mri}
    \vspace{1cm}
    \centering
    \resizebox{\linewidth}{!}{
    \begin{tikzpicture}
    \foreach \i/\m [count = \xi from 0] in {
        0/10.56,
        1/24.12,
        2/28.29,
        10/33.98,
        20/34.08,
        32/34.08
        }
    {
        \node[img,label={[mylabel,yshift=-.3cm]$\M x_{\i} \; \m$}] (i\xi) at (2*\xi,0) {\includegraphics[width=1.9cm]{Files_for_erich/fig_app_mri_path/sol_\i.png}};
        \node[img,label={[mylabel,yshift=-.7cm]below:$\overline{\M M}(\M x_{\i})$}] (i\xi) at (2*\xi,-2) {\includegraphics[width=1.9cm]{Files_for_erich/fig_app_mri_path/mask_\i.png}};
    }

    \begin{scope}[xshift=12cm,yshift=-1.2cm,font=\small]
        \begin{axis}[%
                anchor=north west,
                table/col sep=comma,
                width=6cm,
                height=4cm,
                title=PSNR of $\M x_k$,
                title style=
                {
                  xshift=1cm,
                  yshift=-2cm,
                },
                xlabel={$k$},
                ylabel={},
                grid=both,
                grid style={line width=.1pt, draw=gray!10},
                major grid style={line width=.2pt,draw=gray!50},
                minor tick num=5,
                ymin=20,
            ]
            \addplot[line width=1pt,mark=none] table[x=x,y=psnr] {Files_for_erich/fig_app_mri_path/psnr_mri.csv};
        \end{axis}
    \end{scope}
    \begin{scope}[xshift=12cm,yshift=1.3cm,font=\small]
        \begin{semilogyaxis}[%
                anchor=north west,
                table/col sep=comma,
                width=6cm,
                height=4cm,
                title=$\frac{\norm{\M x_{k+1} -\M x_k}}{\norm{\M x_k}}$,
                title style=
                {
                  xshift=1cm,
                  yshift=-1cm,
                },
                ylabel={},
                ymax=0.1,
                grid=both,
                grid style={line width=.1pt, draw=gray!10},
                major grid style={line width=.2pt,draw=gray!50},
                minor tick num=5,
                xmajorticks=false,
            ]
            \addplot[line width=1pt,mark=none] table[x=x,y=res] {Files_for_erich/fig_app_mri_path/psnr_mri.csv};
        \end{semilogyaxis}
    \end{scope}
    \end{tikzpicture}}
    \caption{Solution and mask path for the 8-fold multi-coil MRI reconstruction for a PD image of knee. }
    \label{fig:mri_path}
\end{figure*}

 We provide visual reconstruction examples obtained with different methods from Table \ref{table:reconstruction_performance_mri} in Figure \ref{fig:compare_mri}.
 We also provide the solution path and the convergence plots for DEAL in Figure \ref{fig:mri_path}. 
\begin{table*}[tbp]
        \centering
        \caption{PSNR values for grayscale deblurring.}%
        \label{table:delur}
        \vspace{.1cm}
        \setlength\tabcolsep{2pt}
        \begin{tabular}{lcccccccccccccccc}
        \toprule
        & \multicolumn{6}{c}{$\sigma = 2.55$}  & \multicolumn{6}{c}{$\sigma = 7.65$} \\
         \midrule
        & \multicolumn{3}{c}{17x17} & \multicolumn{3}{c}{27x27} & \multicolumn{3}{c}{17x17} & \multicolumn{3}{c}{27x27} \\
        & C.man & House & Monarch & C.man & House & Monarch & C.man & House & Monarch & C.man & House & Monarch\\
        \toprule
        EPLL  & 29.18& 32.33 & 27.32 & 27.85 & 28.13 & 22.92 & 24.82 & 28.50 & 23.03 & 24.31 & 26.02 & 20.86 \\
         DEAL (Ours) & \underline{31.72} & \underline{35.20} & \underline{32.77} & \underline{31.64} & \underline{35.03} &  \underline{32.48} & \underline{27.89} &  \underline{32.24} & \underline{28.26} & \underline{27.79} & \underline{32.11} & \underline{28.15}\\
        FDN   & 29.09& 29.75 & 29.13 & 28.78 & 29.29 & 28.60 & 26.18 & 28.01 & 25.86 & 26.13 & 27.41 & 25.39\\
        IRCNN & 31.69& 35.04 & 32.71 & 31.56 & 34.73 & 32.42  & 27.70 & 31.94 & 28.23 & 27.58 & 31.55 & 27.99 \\
        DPIR  & \textbf{32.05} & \textbf{35.82} & \textbf{33.38} & \textbf{31.97} & \textbf{35.52} & \textbf{32.99} & \textbf{28.17} & \textbf{32.79} & \textbf{28.48} & \textbf{27.99} & \textbf{32.87} & \textbf{28.27} \\
        \bottomrule
        \end{tabular}
\end{table*}

\subsection{Convergence Plot for Superresolution \label{app_conv_superees}}
\begin{figure*}
    \centering
    \includegraphics[width=\linewidth]{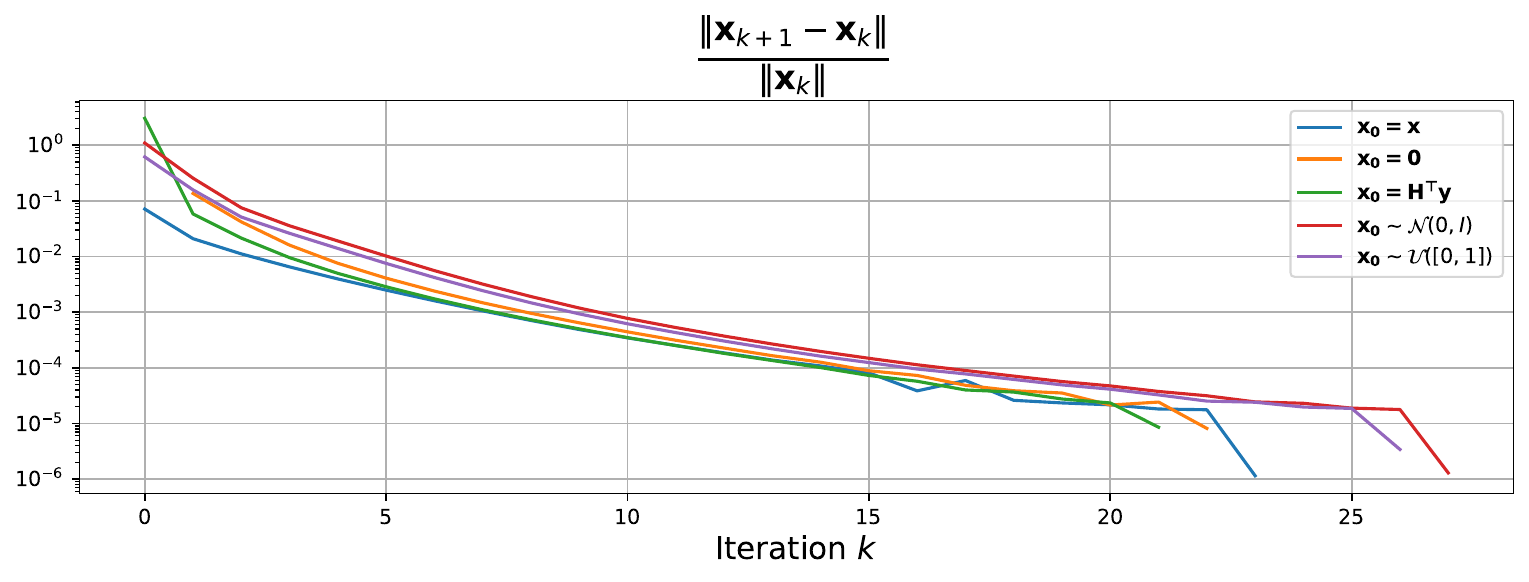}
    \caption{Convergence plot for the super resolution task with $s=2$ and noise level $\sigma_n = 2.55$ for different initializations.} 
    \label{fig:convergence_multires}
\end{figure*}

In Figure \ref{fig:convergence_multires}, we represent the convergence plot for the superresoluion task on the setup of Figure \ref{fig:supperres_compare}.
Again, we empirically observe the convergence of our method.

\section{Visualization of Model Components\label{app_under_the_hood}}

\begin{figure}[tbp]
\centering
\includegraphics[width=1\linewidth]{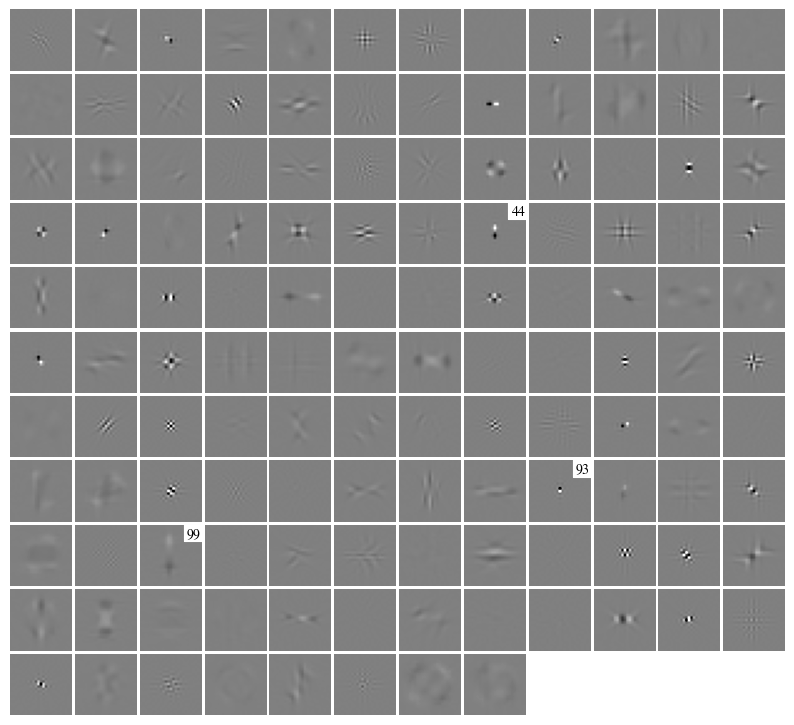}
\vspace{-.1cm}
\caption{Effective convolution kernels for the Multi-Conv block $\M W$ in the grayscale model.
All plots use the same range, where gray corresponds to zero.
Brighter filters are more important.}\label{fig:W_gray_filters}
\end{figure}

\begin{figure*}
    \centering
    \includegraphics[width=1\linewidth]{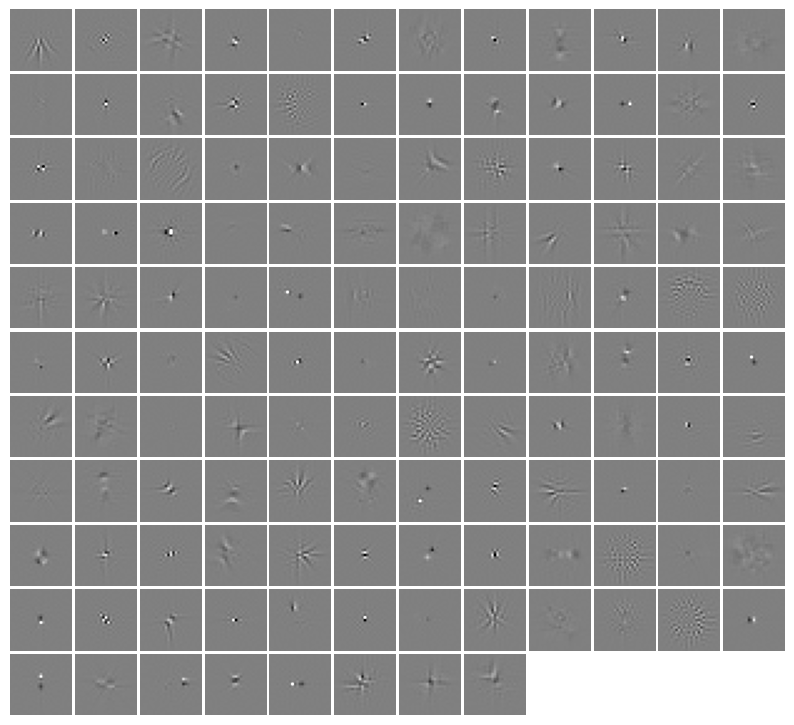}
    \caption{Effective convolution kernels for the Multi-Conv block $\M W_{\mathrm{mask}}$ in the grayscale model.
All plots use the same range, where gray corresponds to zero.
Brighter filters are more important. }
    \label{fig:mask_filters}
\end{figure*}

\begin{figure*}
\centering
\includegraphics[width=\linewidth]{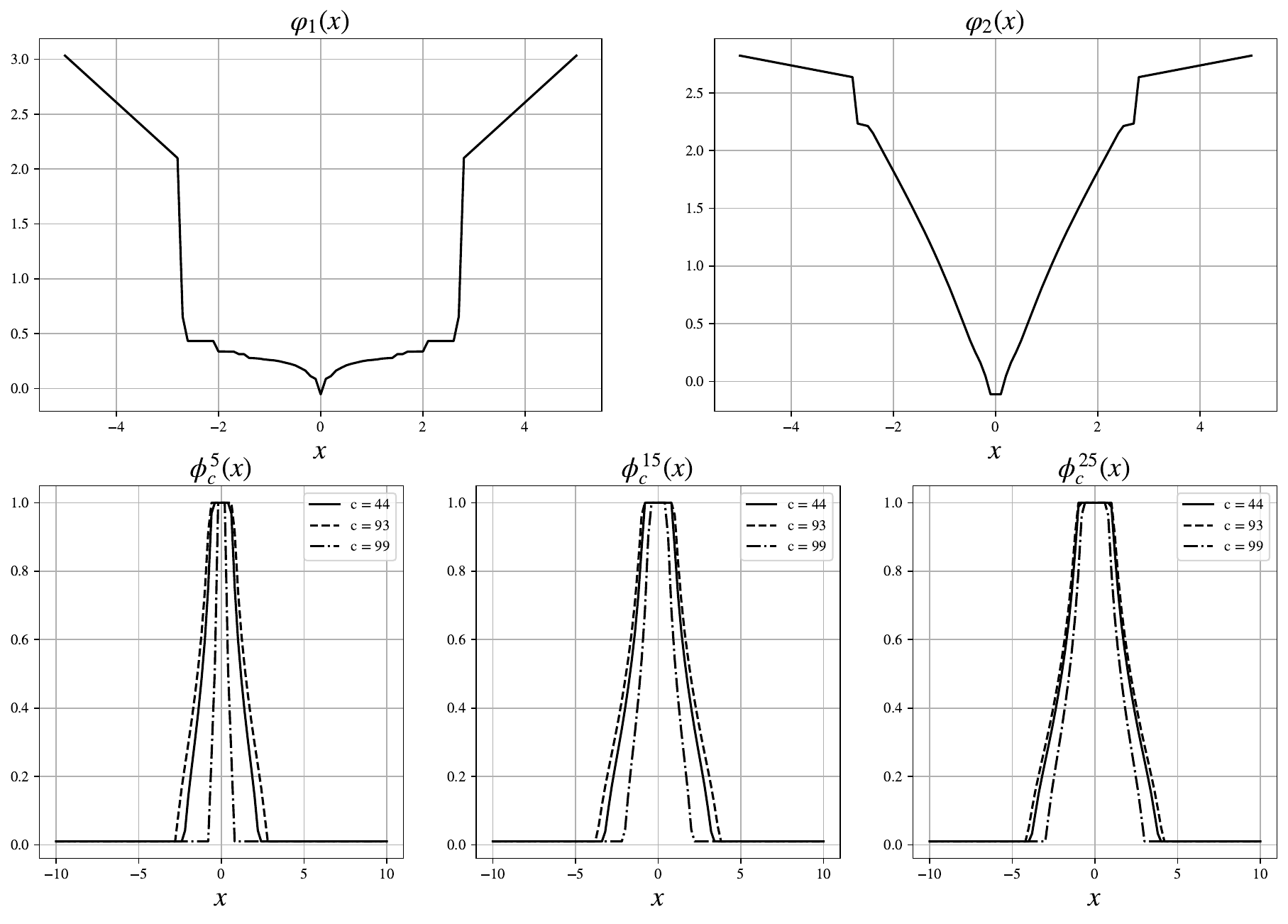}
    \caption{Learned splines in the mask generation network $\M M$ (Figure~\ref{fig:mask_arch}) for grayscale model.}
    \label{fig:splines}
\end{figure*}

\begin{figure*}
    \centering
    \includegraphics[width=\linewidth]{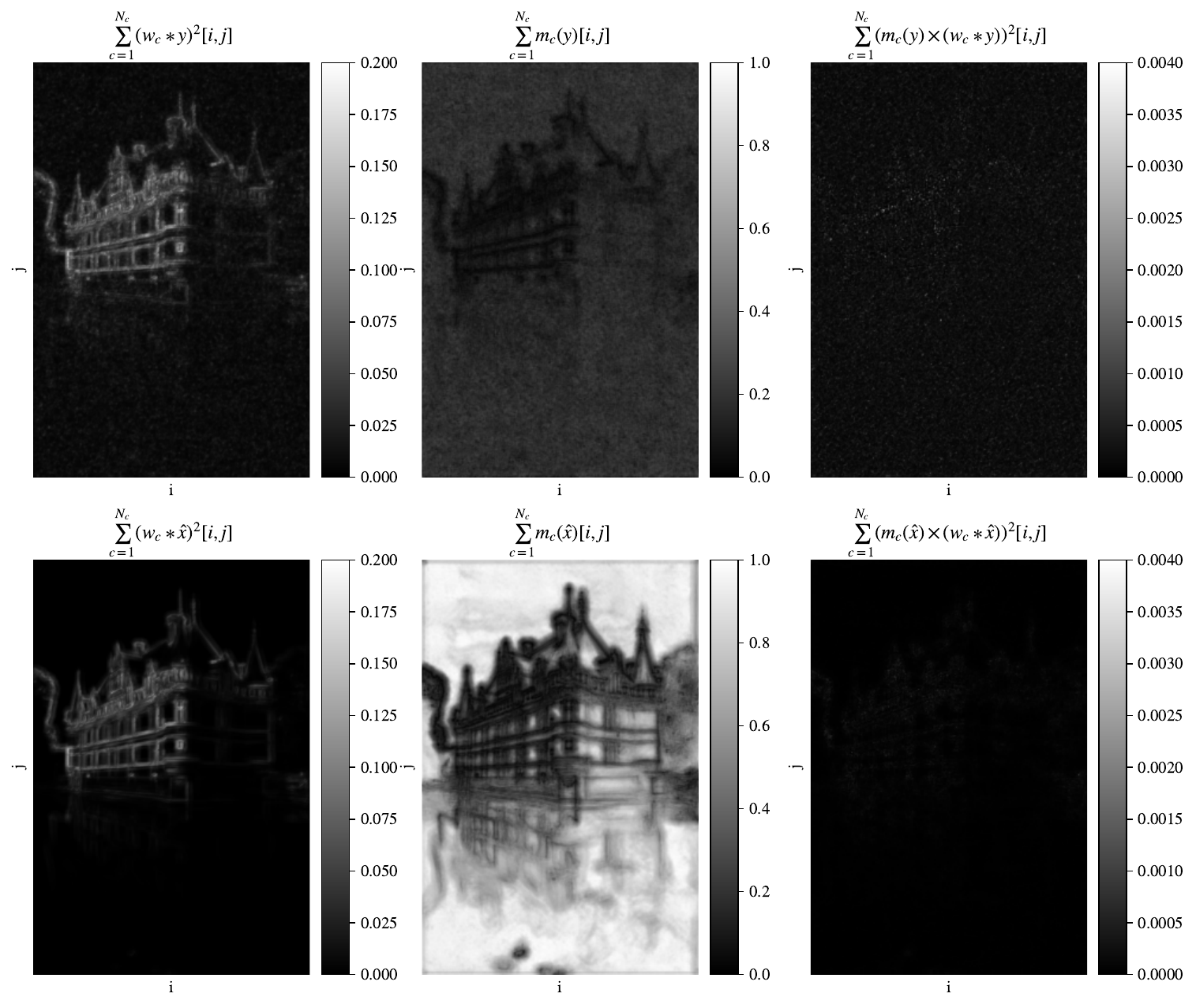}
    \caption{From left to right: a channel-wise average of (i) the squared response to the noisy image (top) and the solution of the castle denoising problem (bottom); (ii) the masks computed on the noisy image (top) and the solution (bottom), (iii) corresponding adapted responses. }
    \label{fig:mask_effect}
\end{figure*}

Now, we inspect the components of our learned grayscale model. In Figure~\ref{fig:W_gray_filters}, we depict the impulse response of $\M W$, i.e., the block's equivalent convolution kernels (a.k.a filters).
The convolutions of $\M W_{\mathrm{mask}}$ have similar structure as we see in Figure \ref{fig:mask_filters}. 
In addition, we depict the learned splines in Figure~\ref{fig:splines}.
For $\phi_c^\sigma$, we visualize three different noise levels $\sigma \in \{5, 15, 25\}$ and channels $c \in \{44, 93, 99\}$. The three visualized channels correspond to vertical edge filters of various scales, see Figure~\ref{fig:W_gray_filters}.
They resemble threshold functions that set high responses to zero. This results in less regularization in the regions that have high responses to the filters that are often activated by the image structures. This is a desirable behavior as the structure of the image should not contribute to the regularization cost.  Additionally, the widths of the last spline $\phi_c^\sigma$ are increasing for all channels with respect to the noise level $\sigma$. Thus, more regularization is performed for higher noise levels. Moreover, we show the channel-wise average of the masks for the noisy and the denoised \emph{castle} image in Figure \ref{fig:mask_effect}. 
The masks remove the undesirable contribution of the image to the filter responses $\M W \M x$ of the regularizer.
This results in lower penalization of the edges and yields sharper solutions.


\end{document}